\documentclass[10pt,journal]{IEEEtran}
\usepackage{times}
\usepackage[pdftex]{graphicx}
\usepackage{amsmath,amssymb,amsthm,bbm}
\usepackage[caption=false,font=footnotesize,labelformat=simple]{subfig}
\usepackage{textcomp}
\usepackage[update,prepend]{epstopdf}
\usepackage{multibib}
\usepackage{url}
\usepackage{color}
\usepackage{multirow}
\usepackage{multicol}
\usepackage{cleveref}
\usepackage{cite}
\usepackage{amsmath,amssymb,amsfonts}
\usepackage{algorithmic}
\usepackage{url}
\def\BibTeX{{\rm B\kern-.05em{\sc i\kern-.025em b}\kern-.08em
		T\kern-.1667em\lower.7ex\hbox{E}\kern-.125emX}}
\IEEEoverridecommandlockouts
\usepackage{etoolbox}
\makeatletter
\patchcmd{\@makecaption}
{\scshape}
{}
{}
{}
\makeatletter
\patchcmd{\@makecaption}
{\\}
{.\ }
{}
{}
\makeatother

\newtheorem{remark}{Remark}

\makeatletter

\newcommand{\comment}[1]{}

\begin{document}
	\date{}
	\title{Deep-Waveform: A Learned OFDM Receiver Based on Deep Complex-valued Convolutional Networks
	}
	

	\author{
		Zhongyuan~Zhao,~\IEEEmembership{Member,~IEEE}, 
		Mehmet~C.~Vuran,~\IEEEmembership{Member,~IEEE}, 
		Fujuan~Guo,~\IEEEmembership{Student Member,~IEEE}, and 
		Stephen~D.~Scott
		\IEEEcompsocitemizethanks{
			\IEEEcompsocthanksitem Zhongyuan~Zhao is with the Department of Electrical and Computer Engineering, Rice University, Houston, Texas 77005, USA (e-mail: zhongyuan.zhao@rice.edu). Mehmet~C.~Vuran, Fujuan Guo, and Stephen~D.~Scott are with the Department of Computer Science and Engineering, University of Nebraska-Lincoln (UNL), Lincoln, Nebraska 68588, USA (e-mail: \{mcvuran;fguo;sscott\}@cse.unl.edu). The majority of this work was completed during the PhD studies of the first author at UNL.
			\IEEEcompsocthanksitem This work is supported by NSF under grant number CNS-1731833.
		}
	}
	
\markboth{Accepted to IEEE Journal on Selected Areas in Communications}{Accepted to IEEE Journal on Selected Areas in Communications}

	\maketitle

	\thispagestyle{empty}
	
\newcommand{\wspace}{\mathbb{W}}				
\newcommand{\bspace}{\mathbb{B}}				
\newcommand{\tspace}{\mathbb{T}}				
\newcommand{\funixt}{_j(c,t)}					
\newcommand{\funxt}{(c,t)}						
\newcommand{\Ang}{\mbox{\rm\AA}}                
\newcommand{\arr}{\ensuremath{\,\rightarrow\,}} 
\newcommand{\Arr}{\ensuremath{\,\Rightarrow\,}} 
\newcommand{\aver}[1]{{<\!\!#1\!\!>}}           
\newcommand{\B}[1]{\ensuremath{\mathbf{#1}}}    
\newcommand{\bra}[1]{\ensuremath{\langle #1|}}  
\newcommand{\bracket}[1]{\ensuremath{\langle #1|#1\rangle}}     
\newcommand{\cket}[1]{\ensuremath{#1\rangle}}   
\newcommand{\const}{\ensuremath{\mathfrak{const}}} 
\newcommand{\D}{\ensuremath{\mathfrak{D}}}      %
\newcommand{\ddotvec}[1]{
    \savebox{\hght}{$\vec{#1}$}\ddot{\raisebox{0pt}[.8\ht\hght]{$\vec{#1}$}}}
\newcommand{\degr}{\ensuremath{^\circ}}         
\newcommand{\diam}{\ensuremath{\varnothing\,}}  
\newcommand{\diver}{\mathop{\mathrm{div}}\nolimits} 
\newcommand{\dotvec}[1]{
    \savebox{\hght}{$\vec{#1}$}\dot{\raisebox{0pt}[.8\ht\hght]{$\vec{#1}$}}}
\newcommand{\dpartder}[2]{\dfrac{\partial^2 #1}{\partial #2^2}} 
\newcommand{\e}{\mathop{\mathrm e}\nolimits}    
\renewcommand{\epsilon}{\varepsilon}            
\newcommand{\frc}[2]{\raisebox{2pt}{$#1$}\big/\raisebox{-3pt}{$#2$}}    
\newcommand{\FT}[1]{\mathcal{F}(#1)}            
\renewcommand{\H}{\ensuremath{\mathfrak{H}}}    %
\newcommand{\IFT}[1]{\mathcal{F}^{-1}(#1)}      
\renewcommand{\ge}{\geqslant}                   
\newcommand{\grad}{\mathop{\mathrm{grad}}\nolimits} 
\newcommand{\I}{\ensuremath{\mathfrak{I}}}      
\newcommand{\indfrac}[2]{\raisebox{2pt}{$\frac{\mbox{\small $#1$}}{\mbox{\small $#2$}}$}} 
\newcommand{\ILT}[1]{\mathop{\mathfrak{L}}\nolimits^{-1}(#1)} 
\newcommand{\Infint}{\int\limits_{-\infty}^\infty} 
\newcommand{\Int}{\int\limits}          
\newcommand{\IInt}{\mathop{{\int\!\!\!\int}}\limits}    
\renewcommand{\kappa}{\varkappa}                
\newcommand{\ket}[1]{\ensuremath{|#1\rangle}}   
\renewcommand{\le}{\leqslant}                   
\newcommand{\ltextarrow}[1]{\ensuremath{\stackrel{#1}\leftarrow}} 
\newcommand{\lvec}{\overrightarrow}             
\newcommand{\LT}[1]{\mathop{\mathfrak{L}}\nolimits(#1)} 
\newcommand{\M}{\ensuremath{\mathop{\mathfrak M}\nolimits}} 
\newcommand{\mean}[1]{\overline{#1}}            
\newcommand{\med}[1]{\mathop{\mathrm{med} #1}\nolimits} 
\newcommand{\Oint}{\oint\limits}                
\renewcommand{\P}{\ensuremath{\mathfrak{P}}}    %
\newcommand{\partder}[2]{\dfrac{\partial #1}{\partial #2}} 
\renewcommand{\phi}{\varphi}                    
\newcommand{\R}{\ensuremath{\mathbb{R}}} 
\newcommand{\rev}[1]{\frac{1}{#1}}              
\newcommand{\rot}{\mathop{\mathrm{rot}}\nolimits} 
\newcommand{\rtextarrow}[1]{\ensuremath{\stackrel{#1}\rightarrow}} 
\newcommand{\Sum}{\sum\limits}  
\newcommand{\sinc}{\mathop{\mathrm{sinc}}\nolimits} 
\newcommand{\Tr}{\mathop{\mathrm{Tr}}\nolimits} 
\newcommand{\veci}{{\vec\imath}}                
\newcommand{\vecj}{{\vec\jmath}}                
\newcommand{\veck}{{\vec{k}}}                   
\newcommand{\when}[2]{\settowidth{\myflt}{\scriptsize $#2$}
    \ensuremath{\left.{#1}\right|_{#2}\hspace{-\myflt}\,}}
\newcommand{\ZT}[1]{\mathop{\mathcal{Z}}\nolimits(#1)} 
\newcommand{\IZT}[1]{\mathop{\mathcal{Z}}\nolimits^{-1}(#1)} 
\newcommand*{\Perm}[2]{{}^{#1}\!P_{#2}}	
\newcommand*{\Comb}[2]{{}^{#1}C_{#2}}	
\newcommand{\erfi}{\mathrm{erfi}}
\newcommand{\diff}{\mathrm{d}}

%

\newenvironment{myproof}[1][$\!\!$]{{\noindent\bf Proof #1: }}
                         {\hfill\QED\medskip}

\newenvironment{mylist}{\begin{list}{}{  \setlength{\itemsep  }{2pt} \setlength{\parsep    }{0in}
                                         \setlength{\parskip  }{0in} \setlength{\topsep    }{5pt}
                                         \setlength{\partopsep}{0in} \setlength{\leftmargin}{2pt}
                                         \setlength{\labelsep }{5pt} \setlength{\labelwidth}{-5pt}}}
                          {\end{list}\medskip}

\newcounter{excercise}
\newcounter{excercisepart}
\newcommand \excercise[1]{\addtocounter{excercise}{1} \setcounter{excercisepart}{0} \medskip
						  \noindent {\bf \theexcercise\ \, #1}}
\newcommand \excercisepart[1]{\addtocounter{excercisepart}{1} \medskip
						      \noindent {\it \Alph{excercisepart}\ \, #1}}


\newenvironment{slideeq} {              \begin{equation*}} {\end{equation*}            }
\newenvironment{nslideeq}{              \begin{equation*}} {\end{equation*}            }
\newenvironment{sslideeq}{\small        \begin{equation*}} {\end{equation*} \normalfont}
\newenvironment{fslideeq}{\footnotesize \begin{equation*}} {\end{equation*} \normalfont}
\newenvironment{slidealign} {              \begin{align*}} {\end{align*}            }
\newenvironment{nslidealign}{              \begin{align*}} {\end{align*}            }
\newenvironment{sslidealign}{\small        \begin{align*}} {\end{align*} \normalfont}
\newenvironment{fslidealign}{\footnotesize \begin{align*}} {\end{align*} \normalfont}

\definecolor{pennblue}{cmyk}{1,0.65,0,0.30}
\definecolor{pennred}{cmyk}{0,1,0.65,0.34}
\definecolor{mygreen}{rgb}{0.10,0.50,0.10}
\newcommand \red[1]         {{\color{red}#1}}
\newcommand \black[1]         {{\color{black}#1}}
\newcommand \blue[1]        {{\color{blue}#1}}
\newcommand \grey[1]        {{\color[rgb]{0.80,0.80,0.80}#1}}
\newcommand \green[1]       {{\color[rgb]{0.10,0.50,0.10}#1}}
\newcommand \bulletcolor[1] {{\color{pennblue}#1}}
\def \arrowbullet {\bulletcolor{$\ \Rightarrow\ $}}
\def \arrbullet   {\bulletcolor{$\ \Rightarrow\ $}}
\def \ab          {\bulletcolor{$\ \Rightarrow\ $}}
\def \arritem     {\item[] \quad \arrowbullet}
\def \ai          {\item[] \quad \arrowbullet}
\def \doublearrow {\bulletcolor{$\ \Leftrightarrow\ $}}
\def \darrbullet  {\bulletcolor{$\ \Leftrightarrow\ $}}

\def \defQfunction 
        {Q(u):=(1/\sqrt{2\pi})\int_u^\infty e^{-u^2/2} du}
\def \intinfty  { \int_{-\infty}^{\infty} }

%
\def \ovP {\overline{P}}
\def \ovl {\overline{l}}
\def \ovbbl {\overline{\bbl}}
\def \ovX {\overline{X}}
\def \ovbbX {\overline{\bbX}}
\def \ovp {\overline{p}}
\def \ovbbp {\overline{\bbp}}
\def \ovr {\overline{r}}
\def \ova {\overline{a}}
\def \ovc {\overline{c}}
\def \ovalpha {\overline{\alpha}}

%
\def \undP {\underline{P}}
\def \undl {\underline{l}}
\def \undbbl {\underline{\bbl}}
\def \undX {\underline{X}}
\def \undbbX {\underline{\bbX}}
\def \undp {\underline{p}}
\def \undbbp {\underline{\bbp}}
\def \undr {\underline{r}}
\def \unda {\underline{a}}
\def \undc {\underline{c}}
\def \undalpha {\underline{\alpha}}

%
\def \undovP     {\underline{\ovP}}
\def \undovX     {\underline{\ovX}}
\def \undovbbX   {\underline{\ovbbX}}
\def \undovp     {\underline{\ovp}}
\def \undovbbp   {\underline{\ovbbp}}
\def \undovr     {\underline{\ovr}}
\def \undova     {\underline{\ova}}
\def \undovc     {\underline{\ovc}}
\def \undovalpha {\underline{\ovalpha}}

\def \SNR     {\text{\normalfont SNR}   }
\def \ap      {\text{\normalfont ap}   }
\def \best    {\text{\normalfont best} }
\def \Co      {\text{\normalfont Co}   }
\def \Cov     {\text{\normalfont Cov}  }
\def \cov     {\text{\normalfont cov}  }
\def \dest    {\text{\normalfont dest} }
\def \diag    {\text{\normalfont diag} }
\def \eig     {\text{\normalfont eig}  }
\def \for     {\text{\normalfont for}  }
\def \forall  {\text{\normalfont for all}  }
\def \forsome {\text{\normalfont for some}  }
\def \ML      {\text{\normalfont ML}   }
\def \MLE     {\text{\normalfont MLE}  }
\def \ml      {\text{\normalfont ml}   }
\def \mse     {\text{\normalfont mse}  }
\def \rank    {\text{\normalfont rank} }
\def \sign    {\text{\normalfont sign} }
\def \tr      {\text{\normalfont tr}   }

\def \dB      {\, \text{\normalfont dB} }
\def \ms      {\, \text{\normalfont m}/ \text{\normalfont s}}
\def \kmh     {\, \text{\normalfont km}/ \text{\normalfont h}}
\def \m       {\, \text{\normalfont m} }
\def \s       {\, \text{\normalfont s} }
\def \sec     {\, \text{\normalfont sec.} }
\def \msec    {\, \text{\normalfont msec.} }
\def \cm      {\, \text{\normalfont cm} }
\def \km      {\, \text{\normalfont km} }
\def \GHz     {\, \text{\normalfont GHz} }
\def \Hz      {\, \text{\normalfont Hz} }
\def \MHZ     {\, \text{\normalfont MHz} }
\def \kHZ     {\, \text{\normalfont kHz} }

\newcommand   \E     [1] {{\mathbb E}\left[#1\right]}
\newcommand   \Ec    [1] {{\mathbb E}\left(#1\right)}
\newcommand   \ind   [1] {{\mathbb I \left\{#1\right\}  } }
\renewcommand \Pr    [1] {\text{\normalfont Pr}  \left[#1\right]}
\newcommand   \Prc   [1] {\text{\normalfont Pr}  \left(#1\right)}
\renewcommand \P     [1] {\text{\normalfont P}   \left[#1\right]}
\newcommand   \Pc    [1] {\text{\normalfont P}   \left(#1\right)}
\newcommand   \Pcbig [1] {\text{\normalfont P}   \big(#1 \big)}
\newcommand   \PcBig [1] {\text{\normalfont P}   \Big(#1 \Big)}
\newcommand   \var   [1] {\text{\normalfont var} \left[#1\right]}
\newcommand   \varc  [1] {\text{\normalfont var} \left(#1\right)}
\renewcommand \Re    [1] {\text{\normalfont Re} \left(#1\right)}
\renewcommand \Im    [1] {\text{\normalfont Im} \left(#1\right)}
\newcommand   \der         [2] {\frac{\partial#1}{\partial#2}}
\newcommand   \inlineder   [2] {\partial#1/\partial#2}

\def \naturals {{\mathbb N}}
\def \reals    {{\mathbb R}}
\def \blog { {\bf \log   } }
\def \given{ {\,\big|\,  } }
\newcommand{\st}{\operatornamewithlimits{s.t.}}
\newcommand{\argmax}{\operatornamewithlimits{argmax}}
\newcommand{\argmin}{\operatornamewithlimits{argmin}}

%
\def\bbarA{{\ensuremath{\bar A}}}
\def\bbarB{{\ensuremath{\bar B}}}
\def\bbarC{{\ensuremath{\bar C}}}
\def\bbarD{{\ensuremath{\bar D}}}
\def\bbarE{{\ensuremath{\bar E}}}
\def\bbarF{{\ensuremath{\bar F}}}
\def\bbarG{{\ensuremath{\bar G}}}
\def\bbarH{{\ensuremath{\bar H}}}
\def\bbarI{{\ensuremath{\bar I}}}
\def\bbarJ{{\ensuremath{\bar J}}}
\def\bbarK{{\ensuremath{\bar K}}}
\def\bbarL{{\ensuremath{\bar L}}}
\def\bbarM{{\ensuremath{\bar M}}}
\def\bbarN{{\ensuremath{\bar N}}}
\def\bbarO{{\ensuremath{\bar O}}}
\def\bbarP{{\ensuremath{\bar P}}}
\def\bbarQ{{\ensuremath{\bar Q}}}
\def\bbarR{{\ensuremath{\bar R}}}
\def\bbarW{{\ensuremath{\bar W}}}
\def\bbarU{{\ensuremath{\bar U}}}
\def\bbarV{{\ensuremath{\bar V}}}
\def\bbarS{{\ensuremath{\bar S}}}
\def\bbarT{{\ensuremath{\bar T}}}
\def\bbarX{{\ensuremath{\bar X}}}
\def\bbarY{{\ensuremath{\bar Y}}}
\def\bbarZ{{\ensuremath{\bar Z}}}
\def\bbara{{\ensuremath{\bar a}}}
\def\bbarb{{\ensuremath{\bar b}}}
\def\bbarc{{\ensuremath{\bar c}}}
\def\bbard{{\ensuremath{\bar d}}}
\def\bbare{{\ensuremath{\bar e}}}
\def\bbarf{{\ensuremath{\bar f}}}
\def\bbarg{{\ensuremath{\bar g}}}
\def\bbarh{{\ensuremath{\bar h}}}
\def\bbari{{\ensuremath{\bar i}}}
\def\bbarj{{\ensuremath{\bar j}}}
\def\bbark{{\ensuremath{\bar k}}}
\def\bbarl{{\ensuremath{\bar l}}}
\def\bbarm{{\ensuremath{\bar m}}}
\def\bbarn{{\ensuremath{\bar n}}}
\def\bbaro{{\ensuremath{\bar o}}}
\def\bbarp{{\ensuremath{\bar p}}}
\def\bbarq{{\ensuremath{\bar q}}}
\def\bbarr{{\ensuremath{\bar r}}}
\def\bbarw{{\ensuremath{\bar w}}}
\def\bbaru{{\ensuremath{\bar u}}}
\def\bbarv{{\ensuremath{\bar v}}}
\def\bbars{{\ensuremath{\bar s}}}
\def\bbart{{\ensuremath{\bar t}}}
\def\bbarx{{\ensuremath{\bar x}}}
\def\bbary{{\ensuremath{\bar y}}}
\def\bbarz{{\ensuremath{\bar z}}}

\def\mbA{{\ensuremath{\mathbb A}}}
\def\mbB{{\ensuremath{\mathbb B}}}
\def\mbC{{\ensuremath{\mathbb C}}}
\def\mbD{{\ensuremath{\mathbb D}}}
\def\mbE{{\ensuremath{\mathbb E}}}
\def\mbF{{\ensuremath{\mathbb F}}}
\def\mbG{{\ensuremath{\mathbb G}}}
\def\mbH{{\ensuremath{\mathbb H}}}
\def\mbI{{\ensuremath{\mathbb I}}}
\def\mbJ{{\ensuremath{\mathbb J}}}
\def\mbK{{\ensuremath{\mathbb K}}}
\def\mbL{{\ensuremath{\mathbb L}}}
\def\mbM{{\ensuremath{\mathbb M}}}
\def\mbN{{\ensuremath{\mathbb N}}}
\def\mbO{{\ensuremath{\mathbb O}}}
\def\mbP{{\ensuremath{\mathbb P}}}
\def\mbQ{{\ensuremath{\mathbb Q}}}
\def\mbR{{\ensuremath{\mathbb R}}}
\def\mbS{{\ensuremath{\mathbb S}}}
\def\mbT{{\ensuremath{\mathbb T}}}
\def\mbU{{\ensuremath{\mathbb U}}}
\def\mbV{{\ensuremath{\mathbb V}}}
\def\mbW{{\ensuremath{\mathbb W}}}
\def\mbX{{\ensuremath{\mathbb X}}}
\def\mbY{{\ensuremath{\mathbb Y}}}
\def\mbZ{{\ensuremath{\mathbb Z}}}
\def\ccalA{{\ensuremath{\mathcal A}}}
\def\ccalB{{\ensuremath{\mathcal B}}}
\def\ccalC{{\ensuremath{\mathcal C}}}
\def\ccalD{{\ensuremath{\mathcal D}}}
\def\ccalE{{\ensuremath{\mathcal E}}}
\def\ccalF{{\ensuremath{\mathcal F}}}
\def\ccalG{{\ensuremath{\mathcal G}}}
\def\ccalH{{\ensuremath{\mathcal H}}}
\def\ccalI{{\ensuremath{\mathcal I}}}
\def\ccalJ{{\ensuremath{\mathcal J}}}
\def\ccalK{{\ensuremath{\mathcal K}}}
\def\ccalL{{\ensuremath{\mathcal L}}}
\def\ccalM{{\ensuremath{\mathcal M}}}
\def\ccalN{{\ensuremath{\mathcal N}}}
\def\ccalO{{\ensuremath{\mathcal O}}}
\def\ccalP{{\ensuremath{\mathcal P}}}
\def\ccalQ{{\ensuremath{\mathcal Q}}}
\def\ccalR{{\ensuremath{\mathcal R}}}
\def\ccalW{{\ensuremath{\mathcal W}}}
\def\ccalU{{\ensuremath{\mathcal U}}}
\def\ccalV{{\ensuremath{\mathcal V}}}
\def\ccalS{{\ensuremath{\mathcal S}}}
\def\ccalT{{\ensuremath{\mathcal T}}}
\def\ccalX{{\ensuremath{\mathcal X}}}
\def\ccalY{{\ensuremath{\mathcal Y}}}
\def\ccalZ{{\ensuremath{\mathcal Z}}}
\def\ccala{{\ensuremath{\mathcal a}}}
\def\ccalb{{\ensuremath{\mathcal b}}}
\def\ccalc{{\ensuremath{\mathcal c}}}
\def\ccald{{\ensuremath{\mathcal d}}}
\def\ccale{{\ensuremath{\mathcal e}}}
\def\ccalf{{\ensuremath{\mathcal f}}}
\def\ccalg{{\ensuremath{\mathcal g}}}
\def\ccalh{{\ensuremath{\mathcal h}}}
\def\ccali{{\ensuremath{\mathcal i}}}
\def\ccalj{{\ensuremath{\mathcal j}}}
\def\ccalk{{\ensuremath{\mathcal k}}}
\def\ccall{{\ensuremath{\mathcal l}}}
\def\ccalm{{\ensuremath{\mathcal m}}}
\def\ccaln{{\ensuremath{\mathcal n}}}
\def\ccalo{{\ensuremath{\mathcal o}}}
\def\ccalp{{\ensuremath{\mathcal p}}}
\def\ccalq{{\ensuremath{\mathcal q}}}
\def\ccalr{{\ensuremath{\mathcal r}}}
\def\ccalw{{\ensuremath{\mathcal w}}}
\def\ccalu{{\ensuremath{\mathcal u}}}
\def\ccalv{{\ensuremath{\mathcal v}}}
\def\ccals{{\ensuremath{\mathcal s}}}
\def\ccalt{{\ensuremath{\mathcal t}}}
\def\ccalx{{\ensuremath{\mathcal x}}}
\def\ccaly{{\ensuremath{\mathcal y}}}
\def\ccalz{{\ensuremath{\mathcal z}}}
\def\ccal0{{\ensuremath{\mathcal 0}}}
%
%
\def\hhatA{{\ensuremath{\hat A}}}
\def\hhatB{{\ensuremath{\hat B}}}
\def\hhatC{{\ensuremath{\hat C}}}
\def\hhatD{{\ensuremath{\hat D}}}
\def\hhatE{{\ensuremath{\hat E}}}
\def\hhatF{{\ensuremath{\hat F}}}
\def\hhatG{{\ensuremath{\hat G}}}
\def\hhatH{{\ensuremath{\hat H}}}
\def\hhatI{{\ensuremath{\hat I}}}
\def\hhatJ{{\ensuremath{\hat J}}}
\def\hhatK{{\ensuremath{\hat K}}}
\def\hhatL{{\ensuremath{\hat L}}}
\def\hhatM{{\ensuremath{\hat M}}}
\def\hhatN{{\ensuremath{\hat N}}}
\def\hhatO{{\ensuremath{\hat O}}}
\def\hhatP{{\ensuremath{\hat P}}}
\def\hhatQ{{\ensuremath{\hat Q}}}
\def\hhatR{{\ensuremath{\hat R}}}
\def\hhatW{{\ensuremath{\hat W}}}
\def\hhatU{{\ensuremath{\hat U}}}
\def\hhatV{{\ensuremath{\hat V}}}
\def\hhatS{{\ensuremath{\hat S}}}
\def\hhatT{{\ensuremath{\hat T}}}
\def\hhatX{{\ensuremath{\hat X}}}
\def\hhatY{{\ensuremath{\hat Y}}}
\def\hhatZ{{\ensuremath{\hat Z}}}
\def\hhata{{\ensuremath{\hat a}}}
\def\hhatb{{\ensuremath{\hat b}}}
\def\hhatc{{\ensuremath{\hat c}}}
\def\hhatd{{\ensuremath{\hat d}}}
\def\hhate{{\ensuremath{\hat e}}}
\def\hhatf{{\ensuremath{\hat f}}}
\def\hhatg{{\ensuremath{\hat g}}}
\def\hhath{{\ensuremath{\hat h}}}
\def\hhati{{\ensuremath{\hat i}}}
\def\hhatj{{\ensuremath{\hat j}}}
\def\hhatk{{\ensuremath{\hat k}}}
\def\hhatl{{\ensuremath{\hat l}}}
\def\hhatm{{\ensuremath{\hat m}}}
\def\hhatn{{\ensuremath{\hat n}}}
\def\hhato{{\ensuremath{\hat o}}}
\def\hhatp{{\ensuremath{\hat p}}}
\def\hhatq{{\ensuremath{\hat q}}}
\def\hhatr{{\ensuremath{\hat r}}}
\def\hhatw{{\ensuremath{\hat w}}}
\def\hhatu{{\ensuremath{\hat u}}}
\def\hhatv{{\ensuremath{\hat v}}}
\def\hhats{{\ensuremath{\hat s}}}
\def\hhatt{{\ensuremath{\hat t}}}
\def\hhatx{{\ensuremath{\hat x}}}
\def\hhaty{{\ensuremath{\hat y}}}
\def\hhatz{{\ensuremath{\hat z}}}
%
%
\def\tdA{{\ensuremath{\tilde A}}}
\def\tdB{{\ensuremath{\tilde B}}}
\def\tdC{{\ensuremath{\tilde C}}}
\def\tdD{{\ensuremath{\tilde D}}}
\def\tdE{{\ensuremath{\tilde E}}}
\def\tdF{{\ensuremath{\tilde F}}}
\def\tdG{{\ensuremath{\tilde G}}}
\def\tdH{{\ensuremath{\tilde H}}}
\def\tdI{{\ensuremath{\tilde I}}}
\def\tdJ{{\ensuremath{\tilde J}}}
\def\tdK{{\ensuremath{\tilde K}}}
\def\tdL{{\ensuremath{\tilde L}}}
\def\tdM{{\ensuremath{\tilde M}}}
\def\tdN{{\ensuremath{\tilde N}}}
\def\tdO{{\ensuremath{\tilde O}}}
\def\tdP{{\ensuremath{\tilde P}}}
\def\tdQ{{\ensuremath{\tilde Q}}}
\def\tdR{{\ensuremath{\tilde R}}}
\def\tdW{{\ensuremath{\tilde W}}}
\def\tdU{{\ensuremath{\tilde U}}}
\def\tdV{{\ensuremath{\tilde V}}}
\def\tdS{{\ensuremath{\tilde S}}}
\def\tdT{{\ensuremath{\tilde T}}}
\def\tdX{{\ensuremath{\tilde X}}}
\def\tdY{{\ensuremath{\tilde Y}}}
\def\tdZ{{\ensuremath{\tilde Z}}}
\def\tda{{\ensuremath{\tilde a}}}
\def\tdb{{\ensuremath{\tilde b}}}
\def\tdc{{\ensuremath{\tilde c}}}
\def\tdd{{\ensuremath{\tilde d}}}
\def\tde{{\ensuremath{\tilde e}}}
\def\tdf{{\ensuremath{\tilde f}}}
\def\tdg{{\ensuremath{\tilde g}}}
\def\tdh{{\ensuremath{\tilde h}}}
\def\tdi{{\ensuremath{\tilde i}}}
\def\tdj{{\ensuremath{\tilde j}}}
\def\tdk{{\ensuremath{\tilde k}}}
\def\tdl{{\ensuremath{\tilde l}}}
\def\tdm{{\ensuremath{\tilde m}}}
\def\tdn{{\ensuremath{\tilde n}}}
\def\tdo{{\ensuremath{\tilde o}}}
\def\tdp{{\ensuremath{\tilde p}}}
\def\tdq{{\ensuremath{\tilde q}}}
\def\tdr{{\ensuremath{\tilde r}}}
\def\tdw{{\ensuremath{\tilde w}}}
\def\tdu{{\ensuremath{\tilde u}}}
\def\tdv{{\ensuremath{\tilde r}}}
\def\tds{{\ensuremath{\tilde s}}}
\def\tdt{{\ensuremath{\tilde t}}}
\def\tdx{{\ensuremath{\tilde x}}}
\def\tdy{{\ensuremath{\tilde y}}}
\def\tdz{{\ensuremath{\tilde z}}}
%
\def\chka{{\ensuremath{\check a}}}
\def\chkb{{\ensuremath{\check b}}}
\def\chkc{{\ensuremath{\check c}}}
\def\chkd{{\ensuremath{\check d}}}
\def\chke{{\ensuremath{\check e}}}
\def\chkf{{\ensuremath{\check f}}}
\def\chkg{{\ensuremath{\check g}}}
\def\chkh{{\ensuremath{\check h}}}
\def\chki{{\ensuremath{\check i}}}
\def\chkj{{\ensuremath{\check j}}}
\def\chkk{{\ensuremath{\check k}}}
\def\chkl{{\ensuremath{\check l}}}
\def\chkm{{\ensuremath{\check m}}}
\def\chkn{{\ensuremath{\check n}}}
\def\chko{{\ensuremath{\check o}}}
\def\chkp{{\ensuremath{\check p}}}
\def\chkq{{\ensuremath{\check q}}}
\def\chkr{{\ensuremath{\check r}}}
\def\chkw{{\ensuremath{\check w}}}
\def\chku{{\ensuremath{\check u}}}
\def\chkv{{\ensuremath{\check v}}}
\def\chks{{\ensuremath{\check s}}}
\def\chkt{{\ensuremath{\check t}}}
\def\chkx{{\ensuremath{\check x}}}
\def\chky{{\ensuremath{\check y}}}
\def\chkz{{\ensuremath{\check z}}}
%
%
\def\bbone{{\ensuremath{\mathbf 1}}}
\def\bbzero{{\ensuremath{\mathbf 0}}}
\def\bbA{{\ensuremath{\mathbf A}}}
\def\bbB{{\ensuremath{\mathbf B}}}
\def\bbC{{\ensuremath{\mathbf C}}}
\def\bbD{{\ensuremath{\mathbf D}}}
\def\bbE{{\ensuremath{\mathbf E}}}
\def\bbF{{\ensuremath{\mathbf F}}}
\def\bbG{{\ensuremath{\mathbf G}}}
\def\bbH{{\ensuremath{\mathbf H}}}
\def\bbI{{\ensuremath{\mathbf I}}}
\def\bbJ{{\ensuremath{\mathbf J}}}
\def\bbK{{\ensuremath{\mathbf K}}}
\def\bbL{{\ensuremath{\mathbf L}}}
\def\bbM{{\ensuremath{\mathbf M}}}
\def\bbN{{\ensuremath{\mathbf N}}}
\def\bbO{{\ensuremath{\mathbf O}}}
\def\bbP{{\ensuremath{\mathbf P}}}
\def\bbQ{{\ensuremath{\mathbf Q}}}
\def\bbR{{\ensuremath{\mathbf R}}}
\def\bbW{{\ensuremath{\mathbf W}}}
\def\bbU{{\ensuremath{\mathbf U}}}
\def\bbV{{\ensuremath{\mathbf V}}}
\def\bbS{{\ensuremath{\mathbf S}}}
\def\bbT{{\ensuremath{\mathbf T}}}
\def\bbX{{\ensuremath{\mathbf X}}}
\def\bbY{{\ensuremath{\mathbf Y}}}
\def\bbZ{{\ensuremath{\mathbf Z}}}
\def\bba{{\ensuremath{\mathbf a}}}
\def\bbb{{\ensuremath{\mathbf b}}}
\def\bbc{{\ensuremath{\mathbf c}}}
\def\bbd{{\ensuremath{\mathbf d}}}
\def\bbe{{\ensuremath{\mathbf e}}}
\def\bbf{{\ensuremath{\mathbf f}}}
\def\bbg{{\ensuremath{\mathbf g}}}
\def\bbh{{\ensuremath{\mathbf h}}}
\def\bbi{{\ensuremath{\mathbf i}}}
\def\bbj{{\ensuremath{\mathbf j}}}
\def\bbk{{\ensuremath{\mathbf k}}}
\def\bbl{{\ensuremath{\mathbf l}}}
\def\bbm{{\ensuremath{\mathbf m}}}
\def\bbn{{\ensuremath{\mathbf n}}}
\def\bbo{{\ensuremath{\mathbf o}}}
\def\bbp{{\ensuremath{\mathbf p}}}
\def\bbq{{\ensuremath{\mathbf q}}}
\def\bbr{{\ensuremath{\mathbf r}}}
\def\bbw{{\ensuremath{\mathbf w}}}
\def\bbu{{\ensuremath{\mathbf u}}}
\def\bbv{{\ensuremath{\mathbf v}}}
\def\bbs{{\ensuremath{\mathbf s}}}
\def\bbt{{\ensuremath{\mathbf t}}}
\def\bbx{{\ensuremath{\mathbf x}}}
\def\bby{{\ensuremath{\mathbf y}}}
\def\bbz{{\ensuremath{\mathbf z}}}
\def\bb0{{\ensuremath{\mathbf 0}}}
%

\def\rmA{{\ensuremath\text{A}}}
\def\rmB{{\ensuremath\text{B}}}
\def\rmC{{\ensuremath\text{C}}}
\def\rmD{{\ensuremath\text{D}}}
\def\rmE{{\ensuremath\text{E}}}
\def\rmF{{\ensuremath\text{F}}}
\def\rmG{{\ensuremath\text{G}}}
\def\rmH{{\ensuremath\text{H}}}
\def\rmI{{\ensuremath\text{I}}}
\def\rmJ{{\ensuremath\text{J}}}
\def\rmK{{\ensuremath\text{K}}}
\def\rmL{{\ensuremath\text{L}}}
\def\rmM{{\ensuremath\text{M}}}
\def\rmN{{\ensuremath\text{N}}}
\def\rmO{{\ensuremath\text{O}}}
\def\rmP{{\ensuremath\text{P}}}
\def\rmQ{{\ensuremath\text{Q}}}
\def\rmR{{\ensuremath\text{R}}}
\def\rmW{{\ensuremath\text{W}}}
\def\rmU{{\ensuremath\text{U}}}
\def\rmV{{\ensuremath\text{V}}}
\def\rmS{{\ensuremath\text{S}}}
\def\rmT{{\ensuremath\text{T}}}
\def\rmX{{\ensuremath\text{X}}}
\def\rmY{{\ensuremath\text{Y}}}
\def\rmZ{{\ensuremath\text{Z}}}
\def\rma{{\ensuremath\text{a}}}
\def\rmb{{\ensuremath\text{b}}}
\def\rmc{{\ensuremath\text{c}}}
\def\rmd{{\ensuremath\text{d}}}
\def\rme{{\ensuremath\text{e}}}
\def\rmf{{\ensuremath\text{f}}}
\def\rmg{{\ensuremath\text{g}}}
\def\rmh{{\ensuremath\text{h}}}
\def\rmi{{\ensuremath\text{i}}}
\def\rmj{{\ensuremath\text{j}}}
\def\rmk{{\ensuremath\text{k}}}
\def\rml{{\ensuremath\text{l}}}
\def\rmm{{\ensuremath\text{m}}}
\def\rmn{{\ensuremath\text{n}}}
\def\rmo{{\ensuremath\text{o}}}
\def\rmp{{\ensuremath\text{p}}}
\def\rmq{{\ensuremath\text{q}}}
\def\rmr{{\ensuremath\text{r}}}
\def\rmw{{\ensuremath\text{w}}}
\def\rmu{{\ensuremath\text{u}}}
\def\rmv{{\ensuremath\text{v}}}
\def\rms{{\ensuremath\text{s}}}
\def\rmt{{\ensuremath\text{t}}}
\def\rmx{{\ensuremath\text{x}}}
\def\rmy{{\ensuremath\text{y}}}
\def\rmz{{\ensuremath\text{z}}}
%

%
\def\barbA{{\bar{\ensuremath{\mathbf A}} }}
\def\barbB{{\bar{\ensuremath{\mathbf B}} }}
\def\barbC{{\bar{\ensuremath{\mathbf C}} }}
\def\barbD{{\bar{\ensuremath{\mathbf D}} }}
\def\barbE{{\bar{\ensuremath{\mathbf E}} }}
\def\barbF{{\bar{\ensuremath{\mathbf F}} }}
\def\barbG{{\bar{\ensuremath{\mathbf G}} }}
\def\barbH{{\bar{\ensuremath{\mathbf H}} }}
\def\barbI{{\bar{\ensuremath{\mathbf I}} }}
\def\barbJ{{\bar{\ensuremath{\mathbf J}} }}
\def\barbK{{\bar{\ensuremath{\mathbf K}} }}
\def\barbL{{\bar{\ensuremath{\mathbf L}} }}
\def\barbM{{\bar{\ensuremath{\mathbf M}} }}
\def\barbN{{\bar{\ensuremath{\mathbf N}} }}
\def\barbO{{\bar{\ensuremath{\mathbf O}} }}
\def\barbP{{\bar{\ensuremath{\mathbf P}} }}
\def\barbQ{{\bar{\ensuremath{\mathbf Q}} }}
\def\barbR{{\bar{\ensuremath{\mathbf R}} }}
\def\barbS{{\bar{\ensuremath{\mathbf S}} }}
\def\barbT{{\bar{\ensuremath{\mathbf T}} }}
\def\barbU{{\bar{\ensuremath{\mathbf U}} }}
\def\barbV{{\bar{\ensuremath{\mathbf V}} }}
\def\barbW{{\bar{\ensuremath{\mathbf W}} }}
\def\barbX{{\overline{\bbX}}}
\def\barbY{{\bar{\ensuremath{\mathbf Y}} }}
\def\barbZ{{\bar{\ensuremath{\mathbf Z}} }}
%
%
\def\barba{{\bar{\ensuremath{\mathbf a}} }}
\def\barbb{{\bar{\ensuremath{\mathbf b}} }}
\def\barbc{{\bar{\ensuremath{\mathbf c}} }}
\def\barbd{{\bar{\ensuremath{\mathbf d}} }}
\def\barbe{{\bar{\ensuremath{\mathbf e}} }}
\def\barbf{{\bar{\ensuremath{\mathbf f}} }}
\def\barbg{{\bar{\ensuremath{\mathbf g}} }}
\def\barbh{{\bar{\ensuremath{\mathbf h}} }}
\def\barbi{{\bar{\ensuremath{\mathbf i}} }}
\def\barbj{{\bar{\ensuremath{\mathbf j}} }}
\def\barbk{{\bar{\ensuremath{\mathbf k}} }}
\def\barbl{{\bar{\ensuremath{\mathbf l}} }}
\def\barbm{{\bar{\ensuremath{\mathbf m}} }}
\def\barbn{{\bar{\ensuremath{\mathbf n}} }}
\def\barbo{{\bar{\ensuremath{\mathbf o}} }}
\def\barbp{{\bar{\ensuremath{\mathbf p}} }}
\def\barbq{{\bar{\ensuremath{\mathbf q}} }}
\def\barbr{{\bar{\ensuremath{\mathbf r}} }}
\def\barbs{{\bar{\ensuremath{\mathbf s}} }}
\def\barbt{{\bar{\ensuremath{\mathbf t}} }}
\def\barbu{{\bar{\ensuremath{\mathbf u}} }}
\def\barbv{{\bar{\ensuremath{\mathbf v}} }}
\def\barbw{{\bar{\ensuremath{\mathbf w}} }}
\def\barbx{{\bar{\ensuremath{\mathbf x}} }}
\def\barby{{\bar{\ensuremath{\mathbf y}} }}
\def\barbz{{\bar{\ensuremath{\mathbf z}} }}
%
%
%
\def\hbA{{\hat{\ensuremath{\mathbf A}} }}
\def\hbB{{\hat{\ensuremath{\mathbf B}} }}
\def\hbC{{\hat{\ensuremath{\mathbf C}} }}
\def\hbD{{\hat{\ensuremath{\mathbf D}} }}
\def\hbE{{\hat{\ensuremath{\mathbf E}} }}
\def\hbF{{\hat{\ensuremath{\mathbf F}} }}
\def\hbG{{\hat{\ensuremath{\mathbf G}} }}
\def\hbH{{\hat{\ensuremath{\mathbf H}} }}
\def\hbI{{\hat{\ensuremath{\mathbf I}} }}
\def\hbJ{{\hat{\ensuremath{\mathbf J}} }}
\def\hbK{{\hat{\ensuremath{\mathbf K}} }}
\def\hbL{{\hat{\ensuremath{\mathbf L}} }}
\def\hbM{{\hat{\ensuremath{\mathbf M}} }}
\def\hbN{{\hat{\ensuremath{\mathbf N}} }}
\def\hbO{{\hat{\ensuremath{\mathbf O}} }}
\def\hbP{{\hat{\ensuremath{\mathbf P}} }}
\def\hbQ{{\hat{\ensuremath{\mathbf Q}} }}
\def\hbR{{\hat{\ensuremath{\mathbf R}} }}
\def\hbS{{\hat{\ensuremath{\mathbf S}} }}
\def\hbT{{\hat{\ensuremath{\mathbf T}} }}
\def\hbU{{\hat{\ensuremath{\mathbf U}} }}
\def\hbV{{\hat{\ensuremath{\mathbf V}} }}
\def\hbW{{\hat{\ensuremath{\mathbf W}} }}
\def\hbX{{\hat{\ensuremath{\mathbf X}} }}
\def\hbY{{\hat{\ensuremath{\mathbf Y}} }}
\def\hbZ{{\hat{\ensuremath{\mathbf Z}} }}
%
%
\def\hba{{\hat{\ensuremath{\mathbf a}} }}
\def\hbb{{\hat{\ensuremath{\mathbf b}} }}
\def\hbc{{\hat{\ensuremath{\mathbf c}} }}
\def\hbd{{\hat{\ensuremath{\mathbf d}} }}
\def\hbe{{\hat{\ensuremath{\mathbf e}} }}
\def\hbf{{\hat{\ensuremath{\mathbf f}} }}
\def\hbg{{\hat{\ensuremath{\mathbf g}} }}
\def\hbh{{\hat{\ensuremath{\mathbf h}} }}
\def\hbi{{\hat{\ensuremath{\mathbf i}} }}
\def\hbj{{\hat{\ensuremath{\mathbf j}} }}
\def\hbk{{\hat{\ensuremath{\mathbf k}} }}
\def\hbl{{\hat{\ensuremath{\mathbf l}} }}
\def\hbm{{\hat{\ensuremath{\mathbf m}} }}
\def\hbn{{\hat{\ensuremath{\mathbf n}} }}
\def\hbo{{\hat{\ensuremath{\mathbf o}} }}
\def\hbp{{\hat{\ensuremath{\mathbf p}} }}
\def\hbq{{\hat{\ensuremath{\mathbf q}} }}
\def\hbr{{\hat{\ensuremath{\mathbf r}} }}
\def\hbs{{\hat{\ensuremath{\mathbf s}} }}
\def\hbt{{\hat{\ensuremath{\mathbf t}} }}
\def\hbu{{\hat{\ensuremath{\mathbf u}} }}
\def\hbv{{\hat{\ensuremath{\mathbf v}} }}
\def\hbw{{\hat{\ensuremath{\mathbf w}} }}
\def\hbx{{\hat{\ensuremath{\mathbf x}} }}
\def\hby{{\hat{\ensuremath{\mathbf y}} }}
\def\hbz{{\hat{\ensuremath{\mathbf z}} }}
%
%
%
\def\tbA{{\tilde{\ensuremath{\mathbf A}} }}
\def\tbB{{\tilde{\ensuremath{\mathbf B}} }}
\def\tbC{{\tilde{\ensuremath{\mathbf C}} }}
\def\tbD{{\tilde{\ensuremath{\mathbf D}} }}
\def\tbE{{\tilde{\ensuremath{\mathbf E}} }}
\def\tbF{{\tilde{\ensuremath{\mathbf F}} }}
\def\tbG{{\tilde{\ensuremath{\mathbf G}} }}
\def\tbH{{\tilde{\ensuremath{\mathbf H}} }}
\def\tbI{{\tilde{\ensuremath{\mathbf I}} }}
\def\tbJ{{\tilde{\ensuremath{\mathbf J}} }}
\def\tbK{{\tilde{\ensuremath{\mathbf K}} }}
\def\tbL{{\tilde{\ensuremath{\mathbf L}} }}
\def\tbM{{\tilde{\ensuremath{\mathbf M}} }}
\def\tbN{{\tilde{\ensuremath{\mathbf N}} }}
\def\tbO{{\tilde{\ensuremath{\mathbf O}} }}
\def\tbP{{\tilde{\ensuremath{\mathbf P}} }}
\def\tbQ{{\tilde{\ensuremath{\mathbf Q}} }}
\def\tbR{{\tilde{\ensuremath{\mathbf R}} }}
\def\tbS{{\tilde{\ensuremath{\mathbf S}} }}
\def\tbT{{\tilde{\ensuremath{\mathbf T}} }}
\def\tbU{{\tilde{\ensuremath{\mathbf U}} }}
\def\tbV{{\tilde{\ensuremath{\mathbf V}} }}
\def\tbW{{\tilde{\ensuremath{\mathbf W}} }}
\def\tbX{{\tilde{\ensuremath{\mathbf X}} }}
\def\tbY{{\tilde{\ensuremath{\mathbf Y}} }}
\def\tbZ{{\tilde{\ensuremath{\mathbf Z}} }}
%
%
\def\tba{{\tilde{\ensuremath{\mathbf a}} }}
\def\tbb{{\tilde{\ensuremath{\mathbf b}} }}
\def\tbc{{\tilde{\ensuremath{\mathbf c}} }}
\def\tbd{{\tilde{\ensuremath{\mathbf d}} }}
\def\tbe{{\tilde{\ensuremath{\mathbf e}} }}
\def\tbf{{\tilde{\ensuremath{\mathbf f}} }}
\def\tbg{{\tilde{\ensuremath{\mathbf g}} }}
\def\tbh{{\tilde{\ensuremath{\mathbf h}} }}
\def\tbi{{\tilde{\ensuremath{\mathbf i}} }}
\def\tbj{{\tilde{\ensuremath{\mathbf j}} }}
\def\tbk{{\tilde{\ensuremath{\mathbf k}} }}
\def\tbl{{\tilde{\ensuremath{\mathbf l}} }}
\def\tbm{{\tilde{\ensuremath{\mathbf m}} }}
\def\tbn{{\tilde{\ensuremath{\mathbf n}} }}
\def\tbo{{\tilde{\ensuremath{\mathbf o}} }}
\def\tbp{{\tilde{\ensuremath{\mathbf p}} }}
\def\tbq{{\tilde{\ensuremath{\mathbf q}} }}
\def\tbr{{\tilde{\ensuremath{\mathbf r}} }}
\def\tbs{{\tilde{\ensuremath{\mathbf s}} }}
\def\tbt{{\tilde{\ensuremath{\mathbf t}} }}
\def\tbu{{\tilde{\ensuremath{\mathbf u}} }}
\def\tbv{{\tilde{\ensuremath{\mathbf v}} }}
\def\tbw{{\tilde{\ensuremath{\mathbf w}} }}
\def\tbx{{\tilde{\ensuremath{\mathbf x}} }}
\def\tby{{\tilde{\ensuremath{\mathbf y}} }}
\def\tbz{{\tilde{\ensuremath{\mathbf z}} }}
%
%
\def\bbcalA{\mbox{\boldmath $\mathcal{A}$}}
\def\bbcalB{\mbox{\boldmath $\mathcal{B}$}}
\def\bbcalC{\mbox{\boldmath $\mathcal{C}$}}
\def\bbcalD{\mbox{\boldmath $\mathcal{D}$}}
\def\bbcalE{\mbox{\boldmath $\mathcal{E}$}}
\def\bbcalF{\mbox{\boldmath $\mathcal{F}$}}
\def\bbcalG{\mbox{\boldmath $\mathcal{G}$}}
\def\bbcalH{\mbox{\boldmath $\mathcal{H}$}}
\def\bbcalI{\mbox{\boldmath $\mathcal{I}$}}
\def\bbcalJ{\mbox{\boldmath $\mathcal{J}$}}
\def\bbcalK{\mbox{\boldmath $\mathcal{K}$}}
\def\bbcalL{\mbox{\boldmath $\mathcal{L}$}}
\def\bbcalM{\mbox{\boldmath $\mathcal{M}$}}
\def\bbcalN{\mbox{\boldmath $\mathcal{N}$}}
\def\bbcalO{\mbox{\boldmath $\mathcal{O}$}}
\def\bbcalP{\mbox{\boldmath $\mathcal{P}$}}
\def\bbcalQ{\mbox{\boldmath $\mathcal{Q}$}}
\def\bbcalR{\mbox{\boldmath $\mathcal{R}$}}
\def\bbcalW{\mbox{\boldmath $\mathcal{W}$}}
\def\bbcalU{\mbox{\boldmath $\mathcal{U}$}}
\def\bbcalV{\mbox{\boldmath $\mathcal{V}$}}
\def\bbcalS{\mbox{\boldmath $\mathcal{S}$}}
\def\bbcalT{\mbox{\boldmath $\mathcal{T}$}}
\def\bbcalX{\mbox{\boldmath $\mathcal{X}$}}
\def\bbcalY{\mbox{\boldmath $\mathcal{Y}$}}
\def\bbcalZ{\mbox{\boldmath $\mathcal{Z}$}}
%
%
%
%
%
%
\def\tdupsilon{\tilde\upsilon}
\def\tdalpha{\tilde\alpha}
\def\tbeta{\tilde\beta}
\def\tdgamma{\tilde\gamma}
\def\tddelta{\tilde\delta}
\def\tdepsilon{\tilde\epsilon}
\def\tdvarepsilon{\tilde\varepsilon}
\def\tdzeta{\tilde\zeta}
\def\tdeta{\tilde\eta}
\def\tdtheta{\tilde\theta}
\def\tdvartheta{\tilde\vartheta}

\def\tdiota{\tilde\iota}
\def\tdkappa{\tilde\kappa}
\def\tdlambda{\tilde\lambda}
\def\tdmu{\tilde\mu}
\def\tdnu{\tilde\nu}
\def\tdxi{\tilde\xi}
\def\tdpi{\tilde\pi}
\def\tdrho{\tilde\rho}
\def\tdvarrho{\tilde\varrho}
\def\tdsigma{\tilde\sigma}
\def\tdvarsigma{\tilde\varsigma}
\def\tdtau{\tilde\tau}
\def\tdupsilon{\tilde\upsilon}
\def\tdphi{\tilde\phi}
\def\tdvarphi{\tilde\varphi}
\def\tdchi{\tilde\chi}
\def\tdpsi{\tilde\psi}
\def\tdomega{\tilde\omega}

\def\tdGamma{\tilde\Gamma}
\def\tdDelta{\tilde\Delta}
\def\tdTheta{\tilde\Theta}
\def\tdLambda{\tilde\Lambda}
\def\tdXi{\tilde\Xi}
\def\tdPi{\tilde\Pi}
\def\tdSigma{\tilde\Sigma}
\def\tdUpsilon{\tilde\Upsilon}
\def\tdPhi{\tilde\Phi}
\def\tdPsi{\tilde\Psi}
%
%
\def\bbarupsilon{\bar\upsilon}
\def\bbaralpha{\bar\alpha}
\def\bbarbeta{\bar\beta}
\def\bbargamma{\bar\gamma}
\def\bbardelta{\bar\delta}
\def\bbarepsilon{\bar\epsilon}
\def\bbarvarepsilon{\bar\varepsilon}
\def\bbarzeta{\bar\zeta}
\def\bbareta{\bar\eta}
\def\bbartheta{\bar\theta}
\def\bbarvartheta{\bar\vartheta}

\def\bbariota{\bar\iota}
\def\bbarkappa{\bar\kappa}
\def\bbarlambda{\bar\lambda}
\def\bbarmu{\bar\mu}
\def\bbarnu{\bar\nu}
\def\bbarxi{\bar\xi}
\def\bbarpi{\bar\pi}
\def\bbarrho{\bar\rho}
\def\bbarvarrho{\bar\varrho}
\def\bbarvarsigma{\bar\varsigma}
\def\bbarphi{\bar\phi}
\def\bbarvarphi{\bar\varphi}
\def\bbarchi{\bar\chi}
\def\bbarpsi{\bar\psi}
\def\bbaromega{\bar\omega}

\def\bbarGamma{\bar\Gamma}
\def\bbarDelta{\bar\Delta}
\def\bbarTheta{\bar\Theta}
\def\bbarLambda{\bar\Lambda}
\def\bbarXi{\bar\Xi}
\def\bbarPi{\bar\Pi}
\def\bbarSigma{\bar\Sigma}
\def\bbarUpsilon{\bar\Upsilon}
\def\bbarPhi{\bar\Phi}
\def\bbarPsi{\bar\Psi}
%
%
%
%
\def\chkupsilon{\check\upsilon}
\def\chkalpha{\check\alpha}
\def\chkbeta{\check\beta}
\def\chkgamma{\check\gamma}
\def\chkdelta{\check\delta}
\def\chkepsilon{\check\epsilon}
\def\chkvarepsilon{\check\varepsilon}
\def\chkzeta{\check\zeta}
\def\chketa{\check\eta}
\def\chktheta{\check\theta}
\def\chkvartheta{\check\vartheta}

\def\chkiota{\check\iota}
\def\chkkappa{\check\kappa}
\def\chklambda{\check\lambda}
\def\chkmu{\check\mu}
\def\chknu{\check\nu}
\def\chkxi{\check\xi}
\def\chkpi{\check\pi}
\def\chkrho{\check\rho}
\def\chkvarrho{\check\varrho}
\def\chksigma{\check\sigma}
\def\chkvarsigma{\check\varsigma}
\def\chktau{\check\tau}
\def\chkupsilon{\check\upsilon}
\def\chkphi{\check\phi}
\def\chkvarphi{\check\varphi}
\def\chkchi{\check\chi}
\def\chkpsi{\check\psi}
\def\chkomega{\check\omega}

\def\chkGamma{\check\Gamma}
\def\chkDelta{\check\Delta}
\def\chkTheta{\check\Theta}
\def\chkLambda{\check\Lambda}
\def\chkXi{\check\Xi}
\def\chkPi{\check\Pi}
\def\chkSigma{\check\Sigma}
\def\chkUpsilon{\check\Upsilon}
\def\chkPhi{\check\Phi}
\def\chkPsi{\check\Psi}
%
%
%
%

\def\bbalpha{\boldsymbol{\alpha}}
\def\bbbeta{\boldsymbol{\beta}}
\def\bbgamma{\boldsymbol{\gamma}}
\def\bbdelta{\boldsymbol{\delta}}
\def\bbepsilon{\boldsymbol{\epsilon}}
\def\bbvarepsilon{\boldsymbol{\varepsilon}}
\def\bbzeta{\boldsymbol{\zeta}}
\def\bbeta{\boldsymbol{\eta}}
\def\bbtheta{\boldsymbol{\theta}}
\def\bbvartheta{\boldsymbol{\vartheta}}
\def \bbtau {\boldsymbol{\tau}}
\def\bbupsilon{\boldsymbol{\upsilon}}
\def\bbiota{\boldsymbol{\iota}}
\def\bbkappa{\boldsymbol{\kappa}}
\def\bblambda{\boldsymbol{\lambda}}
\def\bblam{\boldsymbol{\lambda}}
\def\bbmu{\boldsymbol{\mu}}
\def\bbnu{\boldsymbol{\nu}}
\def\bbxi{\boldsymbol{\xi}}
\def\bbpi{\boldsymbol{\pi}}
\def\bbrho{\boldsymbol{\rho}}
\def\bbvarrho{\boldsymbol{\varrho}}
\def\bbvarsigma{\boldsymbol{\varsigma}}
\def\bbphi{\boldsymbol{\phi}}
\def\bbvarphi{\boldsymbol{\varphi}}
\def\bbchi{\boldsymbol{\chi}}
\def\bbpsi{\boldsymbol{\psi}}
\def\bbomega{\boldsymbol{\omega}}
\def\bbGamma{\boldsymbol{\Gamma}}
\def\bbDelta{\boldsymbol{\Delta}}
\def\bbTheta{\boldsymbol{\Theta}}
\def\bbLambda{\boldsymbol{\Lambda}}
\def\bbXi{\boldsymbol{\Xi}}
\def\bbPi{\boldsymbol{\Pi}}
\def\bbSigma{\boldsymbol{\Sigma}}
\def\bbUpsilon{\boldsymbol{\Upsilon}}
\def\bbPhi{\boldsymbol{\Phi}}
\def\bbPsi{\boldsymbol{\Psi}}

%
%
\def\barbupsilon{\bar\bbupsilon}
\def\barbalpha{\bar\bbalpha}
\def\barbbeta{\bar\bbbeta}
\def\barbgamma{\bar\bbgamma}
\def\barbdelta{\bar\bbdelta}
\def\barbepsilon{\bar\bbepsilon}
\def\barbvarepsilon{\bar\bbvarepsilon}
\def\barbzeta{\bar\bbzeta}
\def\barbeta{\bar\bbeta}
\def\barbtheta{\bar\bbtheta}
\def\barbvartheta{\bar\bbvartheta}

\def\barbiota{\bar\bbiota}
\def\barbkappa{\bar\bbkappa}
\def\barblambda{\bar\bblambda}
\def\barbmu{\bar\bbmu}
\def\barbnu{\bar\bbnu}
\def\barbxi{\bar\bbxi}
\def\barbpi{\bar\bbpi}
\def\barbrho{\bar\bbrho}
\def\barbvarrho{\bar\bbvarrho}
\def\barbvarsigma{\bar\bbvarsigma}
\def\barbphi{\bar\bbphi}
\def\barbvarphi{\bar\bbvarphi}
\def\barbchi{\bar\bbchi}
\def\barbpsi{\bar\bbpsi}
\def\barbomega{\bar\bbomega}

\def\barbGamma{\bar\bbGamma}
\def\barbDelta{\bar\bbDelta}
\def\barbTheta{\bar\bbTheta}
\def\barbLambda{\bar\bbLambda}
\def\barbXi{\bar\bbXi}
\def\barbPi{\bar\bbPi}
\def\barbSigma{\bar\bbSigma}
\def\barbUpsilon{\bar\bbUpsilon}
\def\barbPhi{\bar\bbPhi}
\def\barbPsi{\bar\bbPsi}
%
%
\def\hbupsilon{\hat\bbupsilon}
\def\hbalpha{\hat\bbalpha}
\def\hbbeta{\hat\bbbeta}
\def\hbgamma{\hat\bbgamma}
\def\hbdelta{\hat\bbdelta}
\def\hbepsilon{\hat\bbepsilon}
\def\hbvarepsilon{\hat\bbvarepsilon}
\def\hbzeta{\hat\bbzeta}
\def\hbeta{\hat\bbeta}
\def\hbtheta{\hat\bbtheta}
\def\hbvartheta{\hat\bbvartheta}

\def\hbiota{\hat\bbiota}
\def\hbkappa{\hat\bbkappa}
\def\hblambda{\hat\bblambda}
\def\hbmu{\hat\bbmu}
\def\hbnu{\hat\bbnu}
\def\hbxi{\hat\bbxi}
\def\hbpi{\hat\bbpi}
\def\hbrho{\hat\bbrho}
\def\hbvarrho{\hat\bbvarrho}
\def\hbvarsigma{\hat\bbvarsigma}
\def\hbphi{\hat\bbphi}
\def\hbvarphi{\hat\bbvarphi}
\def\hbchi{\hat\bbchi}
\def\hbpsi{\hat\bbpsi}
\def\hbomega{\hat\bbomega}

\def\hbGamma{\hat\bbGamma}
\def\hbDelta{\hat\bbDelta}
\def\hbTheta{\hat\bbTheta}
\def\hbLambda{\hat\bbLambda}
\def\hbXi{\hat\bbXi}
\def\hbPi{\hat\bbPi}
\def\hbSigma{\hat\bbSigma}
\def\hbUpsilon{\hat\bbUpsilon}
\def\hbPhi{\hat\bbPhi}
\def\hbPsi{\hat\bbPsi}
%
%
\def\tbupsilon{\tilde\bbupsilon}
\def\tbalpha{\tilde\bbalpha}
\def\tbbeta{\tilde\bbbeta}
\def\tbgamma{\tilde\bbgamma}
\def\tbdelta{\tilde\bbdelta}
\def\tbepsilon{\tilde\bbepsilon}
\def\tbvarepsilon{\tilde\bbvarepsilon}
\def\tbzeta{\tilde\bbzeta}
\def\tbeta{\tilde\bbeta}
\def\tbtheta{\tilde\bbtheta}
\def\tbvartheta{\tilde\bbvartheta}

\def\tbiota{\tilde\bbiota}
\def\tbkappa{\tilde\bbkappa}
\def\tblambda{\tilde\bblambda}
\def\tbmu{\tilde\bbmu}
\def\tbnu{\tilde\bbnu}
\def\tbxi{\tilde\bbxi}
\def\tbpi{\tilde\bbpi}
\def\tbrho{\tilde\bbrho}
\def\tbvarrho{\tilde\bbvarrho}
\def\tbvarsigma{\tilde\bbvarsigma}
\def\tbphi{\tilde\bbphi}
\def\tbvarphi{\tilde\bbvarphi}
\def\tbchi{\tilde\bbchi}
\def\tbpsi{\tilde\bbpsi}
\def\tbomega{\tilde\bbomega}

\def\tbGamma{\tilde\bbGamma}
\def\tbDelta{\tilde\bbDelta}
\def\tbTheta{\tilde\bbTheta}
\def\tbLambda{\tilde\bbLambda}
\def\tbXi{\tilde\bbXi}
\def\tbPi{\tilde\bbPi}
\def\tbSigma{\tilde\bbSigma}
\def\tbUpsilon{\tilde\bbUpsilon}
\def\tbPhi{\tilde\bbPhi}
\def\tbPsi{\tilde\bbPsi}
\def \deltat {\triangle t}
\def \eps    {\epsilon}
\def \lam    {\lambda}
\def \bblam  {\bblambda}
\def \Lam    {\Lambda}
\def \bbLam  {\bbLambda}
%
\def\hhattheta{\hat\theta}

	\begin{abstract}
		The (inverse) discrete Fourier transform (DFT/IDFT) is often perceived as essential to  orthogonal frequency-division multiplexing (OFDM) systems. 
		In this paper, a deep complex-valued convolutional network (DCCN) is developed to recover bits from time-domain OFDM signals without relying on any explicit DFT/IDFT. 
		The DCCN can exploit the cyclic prefix (CP) of OFDM waveform for increased SNR by replacing DFT with a learned linear transform, 
		and has the advantage of combining CP-exploitation, channel estimation, and intersymbol interference (ISI) mitigation, with a complexity of $\ccalO(N^2)$. 
		Numerical tests show that the DCCN receiver can outperform the legacy channel estimators based on ideal and approximate linear minimum mean square error (LMMSE) estimation and a conventional CP-enhanced technique in Rayleigh fading channels with various delay spreads and mobility.
		The proposed approach benefits from the expressive nature of complex-valued neural networks, which, however, currently lack support from popular deep learning platforms.
		In response, guidelines of exact and approximate implementations of a complex-valued convolutional layer are provided for the design and analysis of convolutional networks for wireless PHY.
		Furthermore, a suite of novel training techniques are developed to improve the convergence and generalizability of the trained model in fading channels.
		This work demonstrates the capability of deep neural networks in processing OFDM waveforms and the results suggest that the FFT processor in OFDM receivers can be replaced by a hardware AI accelerator.
	\end{abstract}
	\begin{IEEEkeywords}
		Channel Estimation, OFDM, Deep Learning, Physical Layer, Wireless Communications.
	\end{IEEEkeywords}
	
	\section{{Introduction}}\label{sec:intro}
	
\begin{figure}[!t]
		\centering
		\includegraphics[width=0.8\linewidth]{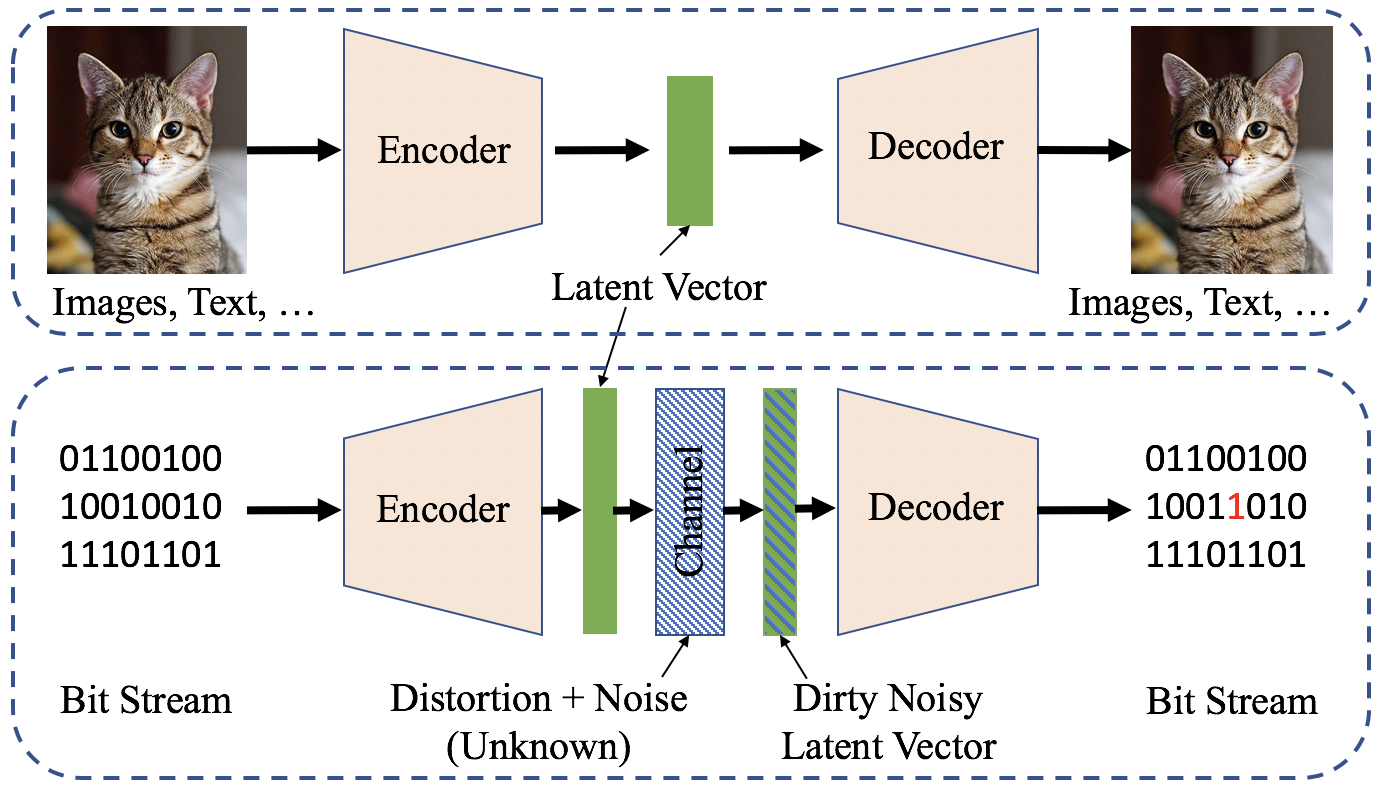}
		\vspace{-0.1in}
		\caption{General AutoEncoder (Top) v.s. Communication AutoEncoder.}
		\vspace{-0.1in}
		\label{fig:sys:ae}
	\end{figure}

Deep learning for the physical layer (PHY) of wireless communications has been explored recently 
for various tasks \cite{TWang17,QMao18,Gunduz2019air}, including signal classification \cite{OShea16c,OShea16b}, parameter estimation \cite{OShea16b,OShea17a,Dorner17,Felix18,HHuang18}, channel estimation \cite{cheng2016channel,HYe18,HHuang18,KYang18,MKim18,Felix18,Liao2019icc,HHe19,Honkala21,lin2020cnn,dong2019cnn,Yang2019doubly,Soltani2019channel}, channel coding \cite{OShea16a,OShea17b,Honkala21}, detection \cite{HYe18,HHe18,fan2019cnn,Zhang2020onebit,Honkala21,Balevi2019onebit}, design of modulation constellation \cite{Dorner17} and pilot design \cite{Xu_2019,mashhadi2020pruning}. Deep neural networks (DNNs) can not only enhance certain functionalities and components of wireless PHY, but also could be developed into an end-to-end novel communication architecture viewed as an autoencoder (AE) \cite{OShea16a,OShea17a,OShea17b,Dorner17,Felix18,lin2020cnn}. 
	Instead of seeking a compact embedding of structured data such as in an image and text (top of Fig.~\ref{fig:sys:ae}), a communication AE (bottom of Fig.~\ref{fig:sys:ae}) generates redundant representation of unstructured bits from which the original information can be recovered after being polluted by a noisy channel. 
	A communication AE learns the behavior of a channel by the structure of data through self-supervised learning of a set of random bits as the training data.
Above the PHY, deep learning is also used in resource allocation and network management, such as traffic prediction \cite{CZhang18}, interference alignment \cite{YHe17}, power control \cite{chowdhury2020unfolding}, spectrum sharing \cite{XLi18}, and scheduling \cite{zhao2020distributed}.

	The data-driven approach of deep learning offers the wireless PHY several advantages: i) achieving synergistic effects of combining cascading modules in the chain of signal processing \cite{OShea17a,Dorner17}, ii) reducing the mismatch between the underlying model and reality, iii) building low complexity solutions by leveraging the non-linearity of DNN \cite{HYe18,Balevi2019onebit,Zhang2020onebit,Honkala21}, and iv) discovering irregular and/or adaptive designs, e.g., for pilot design \cite{Xu_2019,mashhadi2020pruning} and modulation constellation \cite{OShea17a,Dorner17}.
	
	A major distinction of wireless communication compared to image or text interpretation is the reliance on representation of wireless signals in complex field $\mathbb{C}$. 
	However, currently, deep learning for wireless PHY lacks support of complex-valued neural networks (CVNNs) \cite{hirose2012complex,trabelsi2018deep}, which is an emerging area in the machine learning discipline, from popular platforms such TensorFlow \cite{tensorflow2015} and Keras \cite{chollet2015keras}. 
	Instead, existing studies treat the real and imaginary parts of a complex-valued tensor separately,  $\mathbb{C}\Rightarrow\mathbb{R}^2$, such as \textit{approx-D}: two parallel real-valued tensors \cite{Dorner17,HYe18,Honkala21}, or \textit{approx-C}: two channels of a real-valued tensor \cite{mashhadi2020pruning}. 
    However, the complex field $\mathbb{C}$ differs from $\mathbb{R}^2$ space in multiplicative operations, which, 
    if unaccounted for, could prevent the DNN from fully utilizing the relationship between the real and imaginary parts of signal samples in terms of phase and amplitude, resulting in increased complexity, reduced performance, and limited interpretability.
	
	For example,  orthogonal frequency-division multiplexing (OFDM) systems require (inverse) discrete Fourier transform (DFT/IDFT) and/or linear finite impulse-response (FIR) filters that are defined in the complex field $\mathbb{C}$. Without complex-valued representation, existing deep learning-based OFDM receivers \cite{HYe18,HHe18,HHe19,MKim18,Felix18,cheng2016channel,Liao2019icc,Honkala21} and AE \cite{Felix18} are limited to relying on DFT/IDFT in processing the OFDM waveform. Regarding  technical solutions, convolutional neural networks (CNNs) \cite{OShea16b,OShea16c,HHe18,fan2019cnn,mashhadi2020pruning,Honkala21} are less often used than multilayer perceptron (MLP) \cite{OShea16a,OShea17a,OShea17b,Dorner17,Felix18,cheng2016channel,HYe18,KYang18,HHuang18,Zhang2020onebit} in wireless PHY despite their higher efficiency, as CNN depends on the assumptions of the underlying process and is harder to design for the operations in the complex field.
	
	In this paper, we propose a deep complex-valued convolutional network (DCCN) design to recover bits from synchronized time-domain OFDM signals. Instead of relying on DFT/IDFT, the developed end-to-end OFDM receiver learns a new way to receive OFDM waveforms with improved signal-to-noise ratio (SNR), which demonstrates the potential of deep learning for OFDM waveforms.
	The DCCN receiver outperforms the legacy receivers in Rayleigh fading channels with lower complexity by utilizing recent developments within the context of CVNNs. Moreover, many structural and dimensional hyperparameters of the DCCN are selected based on domain knowledge and OFDM frame structure, offering a transferable design template for other waveform structures.
	
	The major contributions of this paper include the following: 
	1) We develop a learned linear transform that can replace DFT/IDFT with increased SNR in processing OFDM waveforms by exploiting its cyclic prefix (CP). The results suggest that a new hardware accelerator can potentially replace the FFT processor in OFDM receivers, and we demonstrate the potential of CVNN in learning communication waveforms. 
	2) We also design a data-driven interpretable DCCN channel equalizer that achieves superior performance than the legacy receivers and good generalizability at a low complexity of $\ccalO(N^2)$ by combining CP exploitation \cite{Quadeer10,Al-Naffouri10, JYang13, Rathinakumar16, LXu17}, intersymbol interference (ISI) mitigation, and channel estimation.
	3) We present a suite of training methods to improve the convergence and generalizability of the DNN-based receiver in fading channels, including a transfer learning scheme that trains the basic OFDM demodulation and channel equalization in two stages. The models are trained and evaluated in different settings of SNR values for white noise and fading channels, and using mixed Rayleigh fading models to smoothen the loss landscape of training.  
	4) To the best of our knowledge, DCCN is the first deep neural network-based wireless receiver that employs explicit complex-valued representation for the entire In-Phase and Quadrature (IQ) domain. A library of complex-valued dense and convolutional layers is provided in open-source software \cite{dlofdm18}, and guidelines for processing complex-valued tensor with real-valued neural networks are provided to the community.

	The rest of this paper is organized as follows: 
	The related work is discussed in Section~\ref{sec:review}.  
	An overview of an OFDM system and channel estimation approaches are provided in Section~\ref{sec:sys}. 
	In Section~\ref{sec:ai}, the design and training approach of the DCCN receiver are presented. 
	The numerical results are presented in Section~\ref{sec:eval}.  
	Finally, the conclusions and future directions are discussed in Section~\ref{sec:dis}.

	\section{Related Work}\label{sec:review}
	
	\subsection{{Deep Neural Networks for Wireless Communications}}\label{sec:review:phy}
	
An 	MLP makes no assumptions about the underlying processes, and is used in channel estimation \cite{cheng2016channel,HYe18,KYang18,HHuang18,Yang2019doubly}, detection \cite{Balevi2019onebit,Zhang2020onebit},  and communication AE \cite{OShea16a,OShea17a,OShea17b,Dorner17,Felix18}. 
	MLP in these studies typically has 3 to 5 layers, and follows two configurations in the signal path.
	The first configuration is \textit{pass-through}: at the receiver (transmitter), MLP takes received data and pilot (input bits) and outputs estimation of the transmitted symbol or soft bits \cite{Dorner17,Felix18,HYe18,Balevi2019onebit,Zhang2020onebit} (transmit IQ samples \cite{OShea17a,OShea17b,Dorner17,Felix18}).
	Additional components may be included for synchronization in AE \cite{OShea17a,OShea17b,Dorner17,Felix18}. 
	The second configuration is the \textit{estimator}: MLP only estimates channel  \cite{cheng2016channel,KYang18,HHuang18,Yang2019doubly} or other parameters \cite{Dorner17,HHuang18} by utilizing the received pilot or signal, and the signal is recovered separately similar to the legacy systems. 
	In this paper, the developed DCCN employs a 4-layer MLP with estimator configuration between convolutional layers for channel estimation. 
	
	CNN is used in signal classification \cite{OShea16c} and recovery \cite{OShea16b}, channel estimation \cite{Koller18ml,mashhadi2020pruning,Honkala21,lin2020cnn,dong2019cnn,Soltani2019channel,Li2020Residual} and detection \cite{fan2019cnn,Honkala21}.
	Although CNNs may not outperform MLP \cite{OShea16a}, it is more efficient and scalable.  
	For example, an OFDM receiver in \cite{Honkala21} matches the linear minimum mean square error (LMMSE) estimator with linear complexity based on a depth-wise separable convolution cascaded in a residual architecture. 
	CNNs are also configured as pass-through \cite{fan2019cnn,Honkala21} and estimator \cite{Koller18ml} in the signal path.
	Other types of DNNs, such as generalized regression neural network (GRNN) \cite{KYang18} and long short-term memory (LSTM) \cite{Liao2019icc,lin2020cnn} are also used in channel estimation, and model-based DNN that unfolds an iterative algorithm \cite{HHe18} is used for detection.
	
	CVNNs are discussed in \cite{hirose2012complex,trabelsi2018deep} but they have not been supported in popular platforms \cite{tensorflow2015,chollet2015keras}. 
	\cite[p.~1]{trabelsi2018deep} indicates that CVNNs "have been marginalized due to the absence of the building blocks".
	In wireless PHY, most CNNs are real-valued \cite{OShea16c,fan2019cnn,mashhadi2020pruning,Honkala21,dong2019cnn,Soltani2019channel,Li2020Residual}, except that a single layer of complex-valued convolution was used in \cite{OShea16b} to achieve amplitude-phase representation, which would require additional phase (un)wrapping in many applications. 
	Based on the analysis in Section~\ref{sec:ai:complex}, the expressive nature of a CVNN can be preserved by an MLP as well as a CNN \cite{mashhadi2020pruning,Honkala21} with proper input format and dimensions, e.g., reception field and number of filters.
	However, the lack of discussions on such settings makes it difficult to scale up the successful design of CNNs \cite{mashhadi2020pruning,Honkala21} with respect to the system parameters including antenna number and DFT size, or transfer these experiences for future work.
	In this work, guidelines for implementing complex-valued convolutional layers are provided and CVNN is used in the entire IQ domain.

	\begin{figure*}[!t]
		\vspace{-0.1in}
		\centering
		\subfloat[]{
			\includegraphics[height=1.9in]{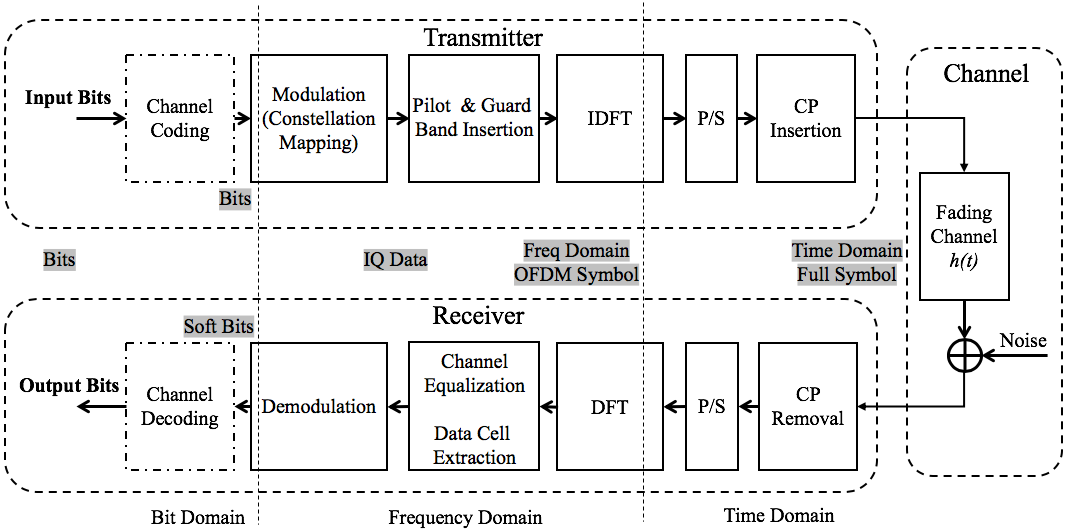}
			\label{fig:ofdm:phy}
		}
		\subfloat[]{
			\includegraphics[height=1.8in]{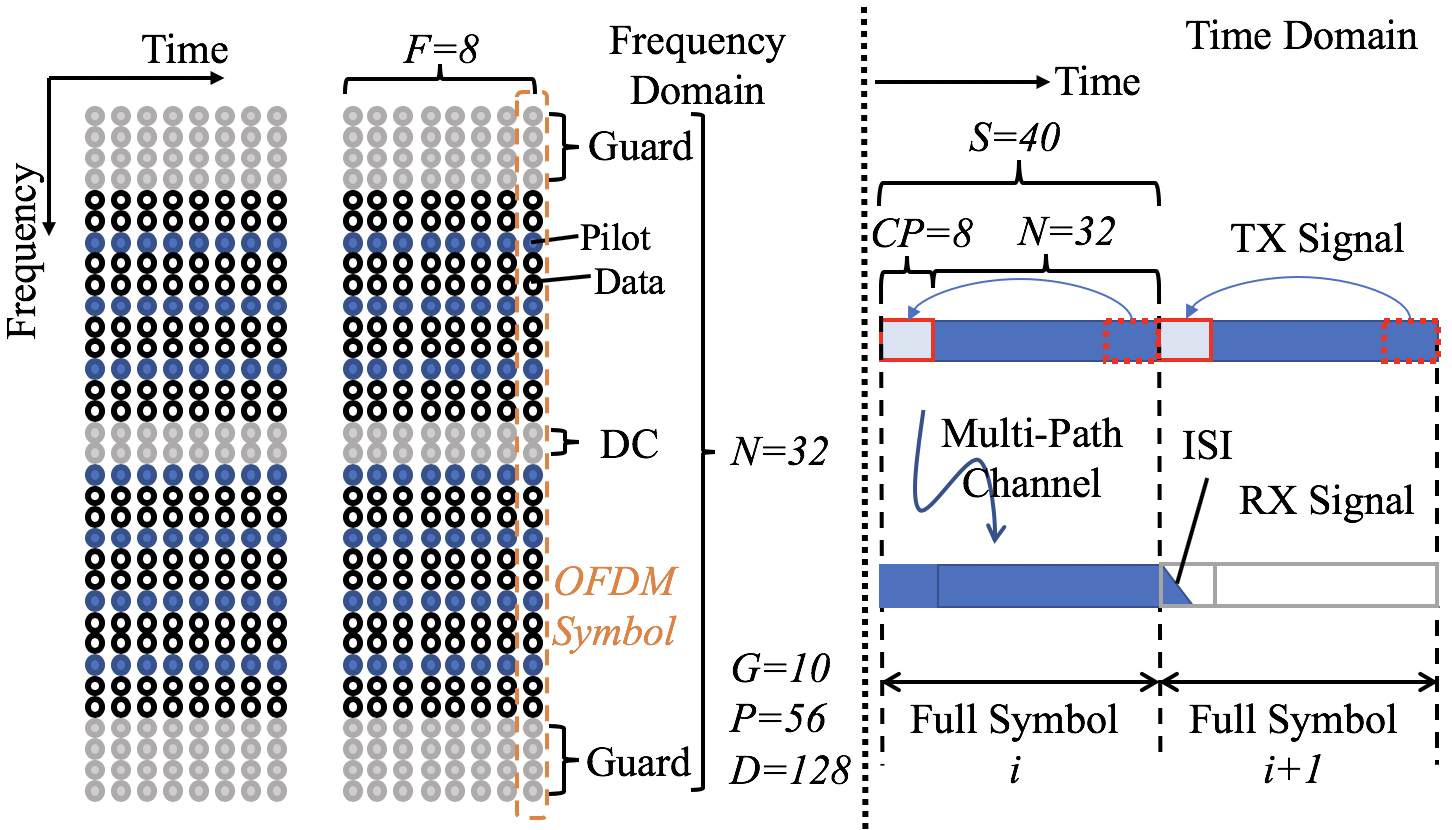}
			\label{fig:ofdm:frame}
		}
		\caption{Physical layer of OFDM system: (a) Diagram of legacy OFDM PHY \cite{shen2006channel}, (b) Exemplary OFDM coherence slot and time-domain waveform \cite{Rathinakumar16}.}
		\vspace{-0.1in}
		\label{fig:ofdm}
	\end{figure*}

	\subsection{OFDM System and Enhanced Channel Estimation}\label{sec:review:ofdm}
	
	OFDM is the most popular system in modern wireless networks. 
	In \cite{JJVanDeBeek98,shen2006channel}, the OFDM physical layer and various channel estimation approaches are introduced. 
	At the transmitter side, several enhanced waveforms are introduced to the OFDM family, such as Filter Bank MultiCarrier (FBMC), UFMC, GFDM for 5G and next generation communication systems \cite{Ankarali17}. 
	These modified OFDM waveforms generally have better characteristics with regards to various interferences. 
	A constellation enhancement approach \cite{WWang18} and deep learning (DL)-based coding system \cite{MKim18} are proposed to reduce the Peak to Average Power Ratio (PAPR) of the OFDM waveform. 
	{An MLP-based channel estimator for FBMC shows good performance in high mobility channels \cite{Cheng2019FBMC}.}
	
	Most of the improvements take place at the receiver side. 
	As a partial copy of the OFDM waveform, cyclic prefix (CP) is introduced for synchronization and ISI mitigation at the receiver, as shown in Fig.~\ref{fig:ofdm:frame}. CP has been exploited to enhance blind \cite{Al-Naffouri10, JYang13, Rathinakumar16, LXu17} and pilot-aided \cite{Quadeer10} channel estimation by improving SNR, frequency selectivity, and interference mitigation \cite{Rathinakumar16}. 
	In \cite{Quadeer10}, the decoded data of a previous OFDM symbol is used to recover the CP of the next symbol. However, the approach in \cite{Quadeer10} can only enhance LS channel estimates. 
	Our model accomplishes this task by processing multiple OFDM symbols simultaneously with better overall performance and slightly improved efficiency without an explicit algorithm, which serves as a complementary to analytical approaches, such as maximum likelihood \cite{Al-Naffouri10,Rathinakumar16} and factor graph \cite{JYang13,LXu17} in exploiting CP.

	DNN is also used to enhance the OFDM receivers. 
	In \cite{HYe18}, a 5-layer MLP is used for channel estimation and symbol detection. 
	Our work confirms the conclusions in \cite{HYe18} that MLP can match MMSE in high SNR and handle adversities related to CP, clipping noise, and channel mismatch better than MMSE. 
	Different from \cite{HYe18}, our channel estimator is linearly activated instead of using ReLU, thus can be interpreted as a low rank approximation of LMMSE. 
	The MLP in \cite{HYe18} handles OFDM waveform without CP, while our model uses CP to simultaneously improve the SNR and mitigate ISI. 
	Moreover, an independent MLP needs to be trained for every $16$ bits in \cite{HYe18}, while we only use one model to process an entire coherence slot, leading to better scalability. 
	{A number of studies use CNN for channel estimation in OFDM systems \cite{Koller18ml,Soltani2019channel,dong2019cnn,Li2020Residual}, without being able to outperform the ideal LMMSE.
	However, for massive MIMO, a low-complexity channel estimator that can outperform approximate LMMSE \cite{dong2019cnn} is still attractive. 
	The OFDM receiver in \cite{Honkala21} achieves linear complexity $\ccalO(N)$, but it only matches the ideal LMMSE on trained channel models and underperforms on unseen channel models.
	Compared to these studies, our DCCN outperforms the ideal LMMSE with a complexity of only $\ccalO(N^2)$ even with unseen channel models. 
	In \cite{Yang2019doubly}, a MLP-based channel estimator for non-OFDM system outperforms approximate LMMSE in doubly selective channels and exhibits better robustness to mobility, of which these findings are confirmed in our work, despite the results are not directly comparable.
	In \cite{Liao2019icc}, long short-term memory (LSTM) is used to estimate channels by predicting future data and is reported to outperform LMMSE in high-speed scenarios. 
	In comparison, our approach mitigates ISI by simultaneously processing multiple OFDM symbols.
	Deep learning has been shown being able to improve MIMO detection \cite{HHe18,fan2019cnn,Zhang2020onebit} and channel coding \cite{lin2020cnn}, while only single antenna is considered in this paper.
	Moreover, the aforementioned work \cite{HYe18,HHe18,HHe19,MKim18,Felix18,cheng2016channel,Liao2019icc,Honkala21} all rely on explicit DFT/IDFT, which is replaced by a learned linear transform in our approach for increased SNR.
	Our work first demonstrates the capability of DNNs in processing OFDM waveform, and our approaches of setting hyperparameters of NN layers based on OFDM frame structure and leveraging domain knowledge offer a transferable design template for other waveform structures.}

	\section{OFDM Communication System}\label{sec:sys}
	We first introduce the relevant concepts and notations in the lower physical layer  of a legacy OFDM system, followed by channel estimation approaches in legacy receivers.

	\subsection{Physical Layer}
	The block diagram of the PHY of an OFDM system is illustrated in Fig.~\ref{fig:ofdm:phy}. 
	At the transmitter, channel encoding is firstly applied to input bits $\bbb\in\{\pm1\}$ for error detection and/or correction. The encoded bits are then converted to complex-valued in-phase and quadrature (IQ) data by mapping to a constellation on the IQ plane. 
	A frequency-domain OFDM symbol, $\bbX$, is created by inserting training signals (pilots) and guard bands into the IQ data, and then $\bbX$ is transformed to a time-domain OFDM symbol, $\bbx$, via an $N$-point IDFT and a subsequent parallel to serial (P/S) conversion. 
	Next, CP, a section of $\bbx$ at its end, is prepended to $\bbx$ to create a time-domain full OFDM symbol, $\bbx_{cp}$,  as illustrated in Fig.~\ref{fig:ofdm:frame}. 
	The baseband signal, $\bbx_{cp}$, is then up-converted to the radio frequency (RF) and transmitted over-the-air by the RF frontend. 
	The radio signal propagates over the wireless channel and is picked and down-converted to baseband IQ samples by the receiver frontend. 
	At the receiver, the received time-domain OFDM symbols, $\bby_{cp}$, are recovered by a carrier synchronizer. 
	Then, the CP is removed from $\bby_{cp}$ and the rest of the IQ samples, $\bby$, are transformed to the frequency-domain OFDM symbol, $\bbY$, via DFT. 
	Based on $\bbY$, a channel equalizer outputs the estimated transmit frequency-domain IQ data $\hat{\bbX}$, which is then demodulated to soft bits (log-likelihood) $\tilde{\bbb}$ and converted to binary output bits $\hat{\bbb}$ by a channel decoder. 
	Finally, $\hat{\bbb}$  is passed to the next layer. Note that the focus of this paper is lower PHY, and channel coding is out of the scope. 
	{We refer the frequency and time domains in Fig.~\ref{fig:ofdm:phy} as the \textit{IQ domain} in which a signal is represented by complex-valued samples.}
	
	OFDM systems typically have a frame structure where {a coherence slot (or ``slot'' for short)} is composed of multiple OFDM symbols, as shown in Fig.~\ref{fig:ofdm:frame}. 
	The notations related to the OFDM coherence slot are as follows: an OFDM symbol contains $N$ subcarriers, where $N$ is the size of DFT/IDFT. 
	Among the $N$ subcarriers, a total of $G$ nullified guard subcarriers are placed at the center (DC guard band) and the edge (edge guard band). 
	A subcarrier in an OFDM symbol is refereed to as a resource element (RE). 
	A coherence slot contains  $F$ consecutive OFDM symbols, in which $P$ and $D$ REs are allocated to pilot and data, respectively. 
	The length of a time-domain full OFDM symbol is $S=N+N_{cp}$ where $N_{cp}$ is the length of CP. Under $m$-ary modulation, an IQ sample carries $m$ bits, and the size of the constellation is $2^m$.

	\subsection{Wireless Channel}

	\begin{figure}[!t]
		\centering
		\vspace{-0.2in} 
		\subfloat[AWGN]{	
			\includegraphics[width=0.31\linewidth]{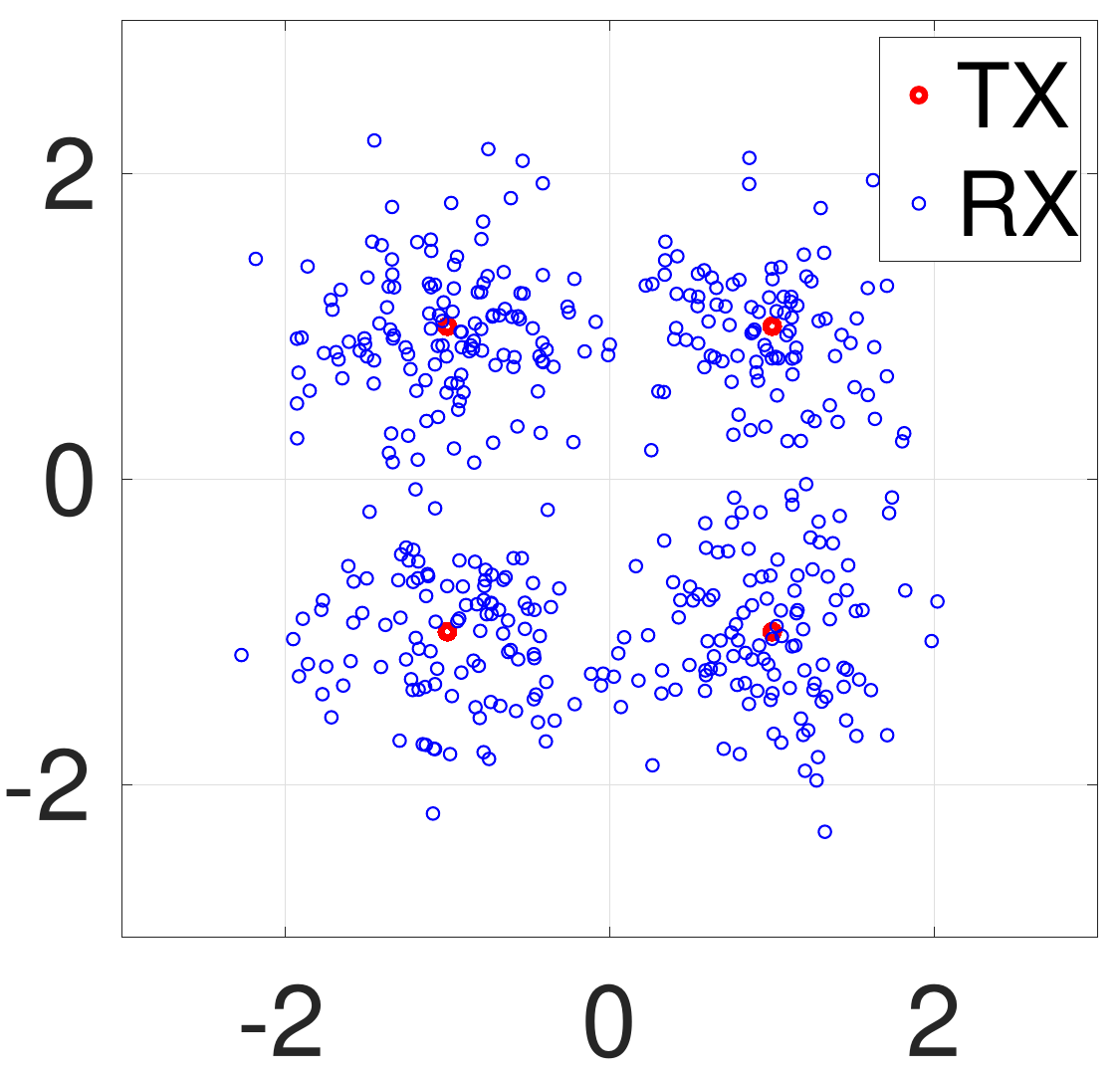}
			\label{fig:channel:awgn}
		}\hspace{-0.15in}
		\subfloat[Flat Fading]{	
			\includegraphics[width=0.31\linewidth]{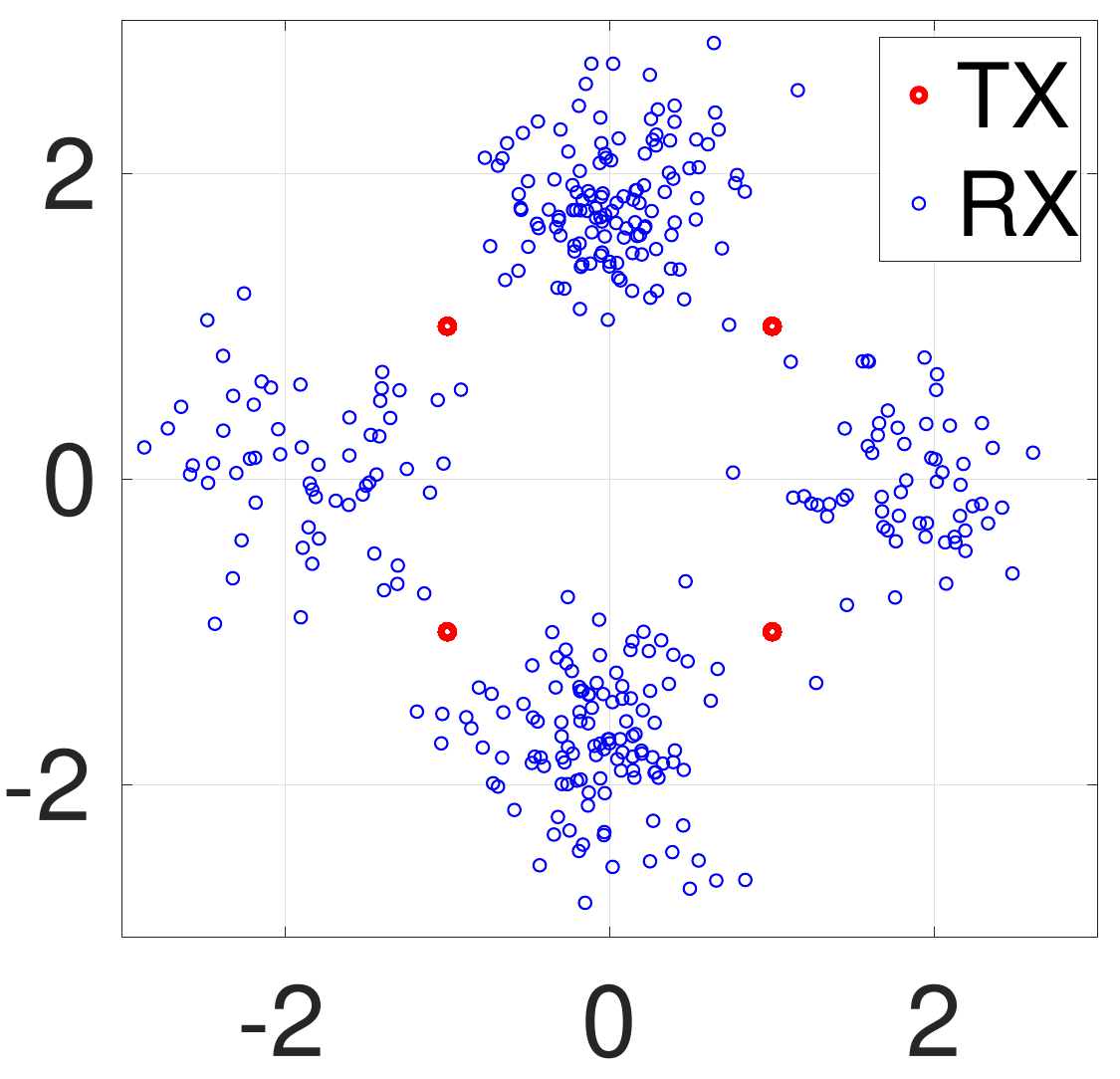}
			\label{fig:channel:flat}
		}\hspace{-0.15in}
		\subfloat[Multipath Fading]{	
			\includegraphics[width=0.31\linewidth]{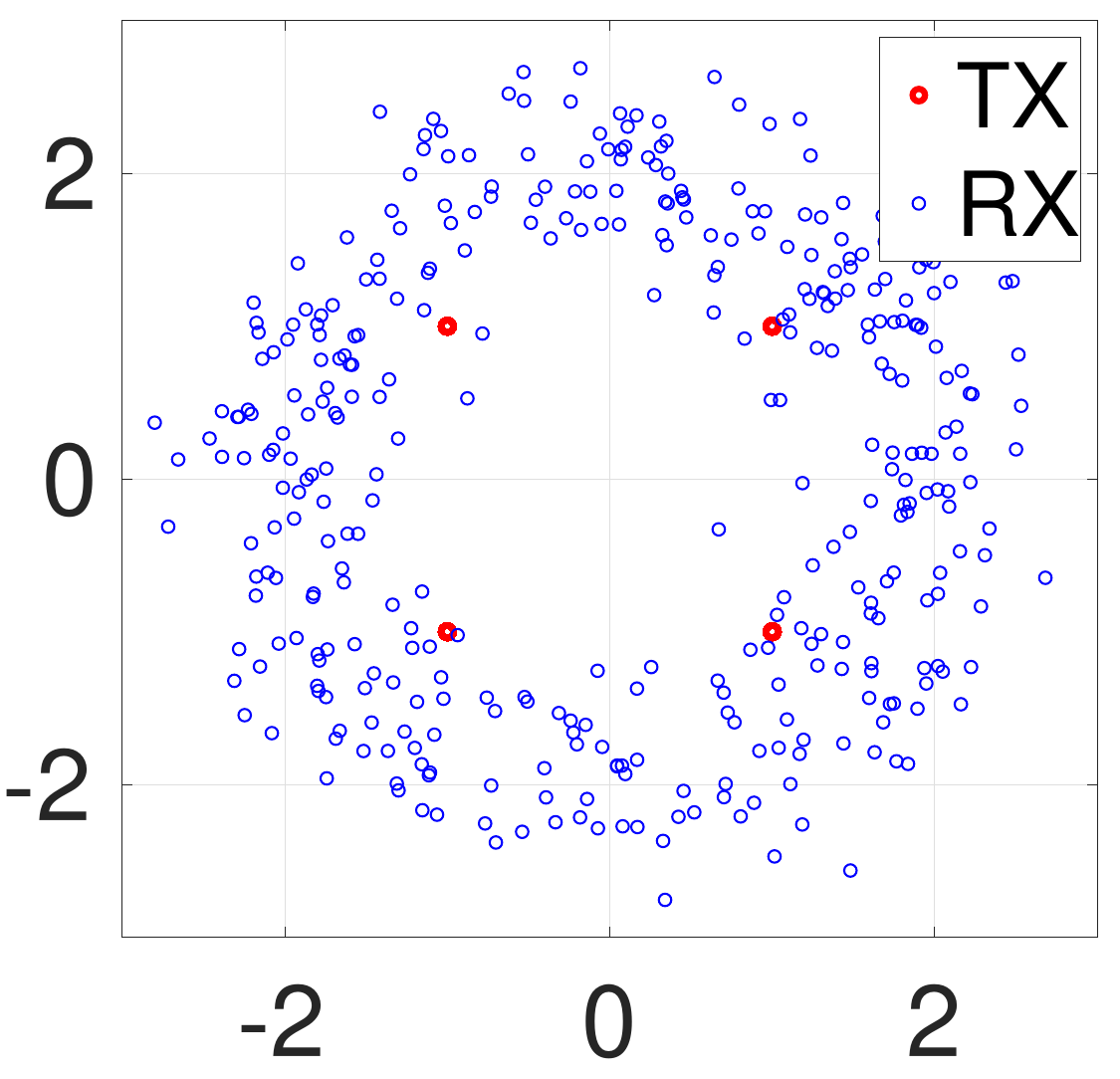}
			\label{fig:channel:EPA}
		}
		\caption{The effects of different wireless channels on QPSK modulation, SNR$=6$~dB, on IQ plane (x: In-Phase, y: Quadrature).}\vspace{-0.1in}
		\label{fig:channel}
	\end{figure}

	A well accepted simplification of the wireless channel model describes the fading and noise processes as \cite{shen2006channel}: 
	\begin{equation}\label{eq:channel:time}
		\bby = \bbx*\bbh + \bbn_o\;,\; \bbY = \bbX\odot \bbH + \bbN_o\;,
	\end{equation}
	{where $\bbx,\bby,\bbn_o \in\mathbb{C}^{SF\times 1}$ are the time-domain transmitted and received signals and white noise, respectively, $\bbh\in\mathbb{C}^{L\times 1}$ is the channel impulse response,  $\bbX,\bbY,\bbH,\bbN_o\in\mathbb{C}^{N\times F}$ are the frequency domain transformations of $\bbx$, $\bby$, $\bbh$, and $\bbn_o$, e.g., $\bbX = DFT_{N}(\bbx)$, and $*$ and $\odot$ are operators of convolution and element-wise product, respectively. Without loss of generality, \eqref{eq:channel:time} can also refer to an OFDM symbol, i.e., $\bbx\in\mathbb{C}^{N\times 1}, \bbx_{cp}\in\mathbb{C}^{S\times 1}$ and $\bbX\in\mathbb{C}^{N\times 1}$.}
	
	The multipath fading can be modeled as a linear finite impulse-response (FIR) filter \cite{jeruchim2006simulation}:
	\begin{equation}\label{eq:channel:tap}
		y_t = \sum_{l=0}^{L-1} x_{t-l}h_l\;,\text{and }h_l=\sum_{k=1}^{K} \sqrt{\Omega_k}z_k\mathrm{sinc}\left(\frac{\tau_k}{T_s}-l\right)\;,
	\end{equation}
	where $z_k$ is a complex-valued random variable, vectors $\mathbf{\Omega}$ and $\bbtau$ represent the power-delay profile (PDP) of the fading process, and $T_s$ is the sampling period of the discrete signal. 
	The filter length $L$ is chosen so that $|h_l|$ is small when $l<0$ or $l\geq L$. 
	For Rayleigh fading, the real and imaginary parts of $z_k$ are i.i.d. Gaussian random variables, thus $|z_k|^2$ follows a Rayleigh distribution. $K$ is the number of paths in a multipath fading channel. 
	In a flat fading channel $K=1$, the channel coefficients on all the subcarriers of an OFDM symbol are identical. 
	In a multipath fading channel, $K>1$, the channel coefficient varies by subcarrier so that the channel exhibits frequency selectivity. 
	The effects of different fading processes on frequency-domain IQ samples are illustrated in Fig.~\ref{fig:channel}. 
	Note that only noise and fading are considered, while channel impairments for channel coding are left for future work. 
	
	{The channel impulse response, $\bbh$, is time-variant, and its coherence time, $T_c$, is inversely proportional to the maximum Doppler frequency $F_d$, i.e.,  $T_c\approx 1/F_d$.
	An OFDM system is usually configured according to its applications, measured coherence time, $T_c$, coherence bandwidth, $B_c$, and total bandwidth, so that slow fading holds for a coherence slot.} 
	
	\begin{figure}[!t]
		\centering
		\includegraphics[width=0.9\linewidth]{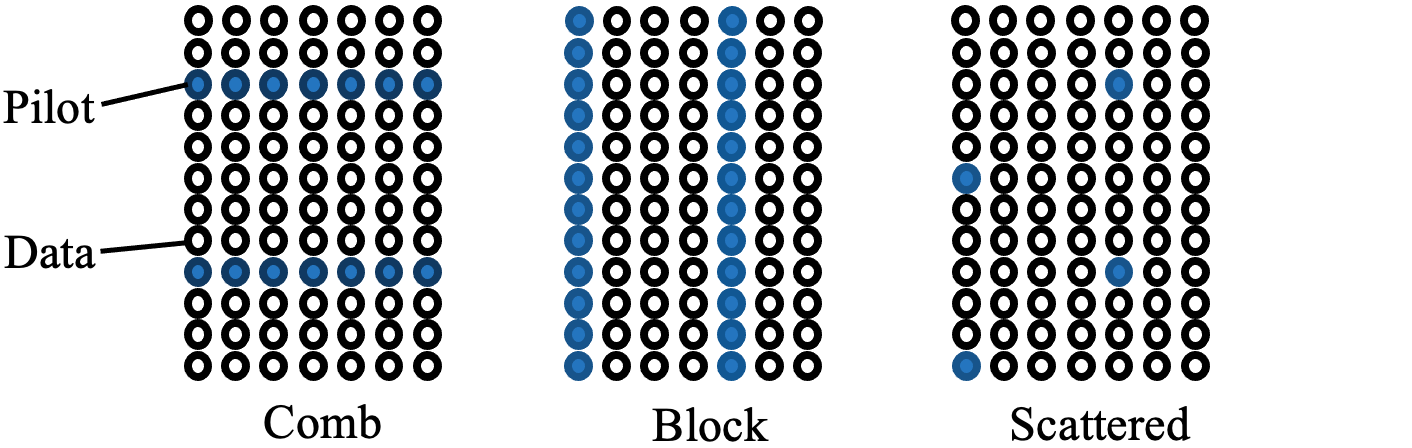}
		\vspace{-0.1in}
		\caption{Typical OFDM pilot patterns: comb, block, and scattered.}
		\vspace{-0.1in}
		\label{fig:ofdm:pilot}
	\end{figure}
	
	\begin{figure*}[!t]
		\vspace{-0.1in}
		\centering
		\includegraphics[width=\linewidth]{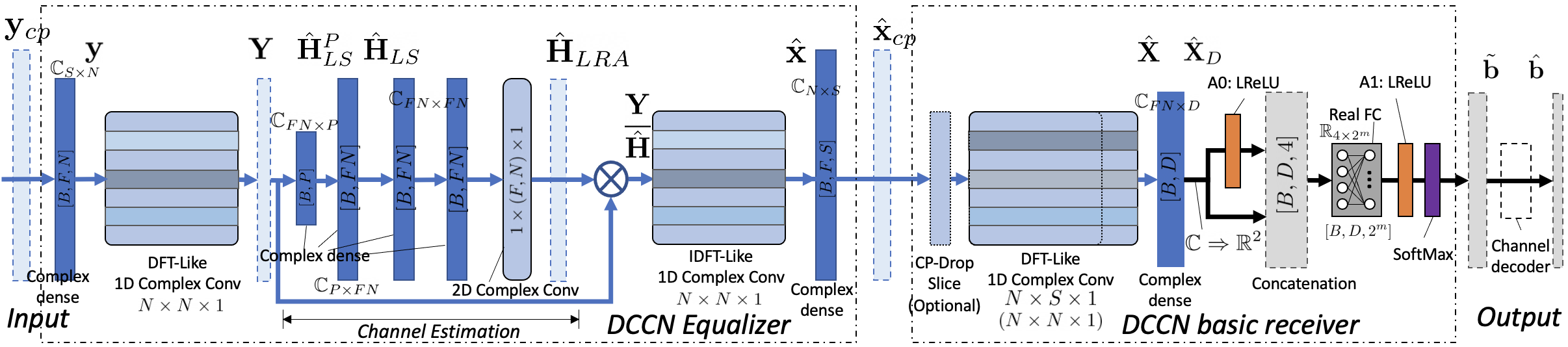}
		\vspace{-0.2in}
		\caption{DCCN OFDM receiver flow graph: a DCCN equalizer converts time domain received signal, $\bby_{cp}$, to estimated transmitted signal, $\bbx_{cp}$, which is subsequently converted to estimated soft bits, $\tilde{\bbb}$, and hard bits, $\hat{\bbb}$, by a DCCN basic receiver. 
		The blue color represents complex-valued domain, and gray for real-valued domain. The dimensions and data types of a dense layer are shown as e.g., $\mathbb{C}_{S\times N}$, and the shape of its output tensor as, e.g., $[B, F, N]$, the dimensions of a Conv layer is labeled beside the block, e.g., $N\times N\times 1$ represents $N$ 1-D filters each of length $N$. A dashed block represents a data tensor. }
		\vspace{-0.1in}
		\label{fig:flowgraph}
	\end{figure*}
	
	\subsection{Channel Estimation and Equalization}\label{sec:sys:equal}
	
	Consider a communication system with a pilot-aided channel estimation at the receiver. Different pilot patterns, such as block, comb, and scattered pilots, as illustrated in Fig.~\ref{fig:ofdm:pilot} \cite{shen2006channel}, are designed to sample the channel distortion. 
	Pilot signals are of either constant signal or low auto-correlation sequence (e.g., Zadoff-Chu sequence) known at the receiver. 
	The basic pilot-aided channel equalizer in OFDM system is based on the least square (LS) estimator \cite{JJVanDeBeek98,shen2006channel}:
	\begin{equation}\label{eq:ls}
		\hat{\bbX} = \frac{\bbY}{\hat{\bbH}_{LS}}\;,\text{ where }\hat{\bbH}_{LS}=\ccalF\left(\frac{\bbY_{P}}{\bbX_{P}}\right)\;, 
	\end{equation}
	where $\hat{\bbX}$ is the estimated signal, $\hat{\mathbf{H}}_{LS}$ contains the LS channel estimates, $\bbX_{P}$ and $\bbY_{P}$ are transmitted and received pilots, respectively, and $\ccalF(\cdot)$ is an interpolation operator, e.g., linear, spline, low-pass-filter, and DFT \cite{JJVanDeBeek98,shen2006channel}. 
	The LS estimator is agnostic to channel statistics, while other estimators, such as LMMSE, maximal likelihood, and parametric channel modeling-based (PCMB) estimator, are based on LS estimation and/or prior channel knowledge \cite{JJVanDeBeek98,shen2006channel}. 
	
	The ideal LMMSE estimator is expressed as \cite{JJVanDeBeek98,Zhou09lmmse}:
	\begin{equation}\label{eq:lmmse}
		\hat{\bbH}_{LMMSE}=\bbR_{\bbH\bbH}\left(\bbR_{\bbH\bbH}+\frac{\beta}{\alpha}\mathbf{I}\right)^{-1}\hat{\mathbf{H}}_{LS}\;,
	\end{equation}
	{where $\bbR_{\bbH\bbH}=\mathrm{E}\{\bbH_{*,i}^{}\bbH_{*,i}^H\}$ is the frequency-domain covariance matrix of channel realization $\bbH$}, $\alpha$ is  the linear-domain SNR, and $\beta$ is a constant defined based on a specific modulation scheme. 
	The ideal LMMSE estimator in \eqref{eq:lmmse} requires prior channel knowledge  $\bbR_{\bbH\bbH}$ and $\alpha$, which are unavailable in practice. Moreover, the matrix inversion in \eqref{eq:lmmse} leads to a computational complexity of $\ccalO(N^3)$. 
	Low-rank approximation (LRA) of LMMSE \cite{edfors1998ofdm} approximates the ideal LMMSE matrix $\bbR_{\bbH\bbH}(\bbR_{\bbH\bbH}+\frac{\beta}{\alpha}\mathbf{I})^{-1}$ via prescribed $\alpha$ and singular value decomposition (SVD) of $\bbR_{\bbH\bbH}$, which can tolerate PDP mismatch and achieve a lower complexity of $\ccalO(N^2)$ at the cost of an irreducible error floor. 
	Improvements on approximate LMMSE (ALMMSE) include using different PDPs \cite{hung10pdp}, SNR estimation \cite{Zhou09lmmse,savaux2015joint}, rank estimation and exploiting the circulant property of LMMSE matrix \cite{Zhou09lmmse}. 
	Fast ALMMSE can further reduce the complexity to $\ccalO(N)$ with a pre-computed LMMSE matrix at the cost of degraded performance in a certain SNR range \cite{ohno2012low}.

	\section{DCCN-Based OFDM Receiver}\label{sec:ai}
	
	The DCCN-based OFDM receiver is denoted by a function $\tilde{\bbb} = \ccalD_{\ccalS}(\bby_{cp};\bbTheta)$, which is defined on a collection of hyperparameters, $\ccalS$, for the configurations of an OFDM frame and DCCN flow-graph (Fig.~\ref{fig:flowgraph}), where $\tilde{\bbb}$ is the log-likelihood of the transmitted bits, $\bby_{cp}$ is the synchronized time-domain received OFDM symbols of a coherence slot including CP, and $\bbTheta$ is the collection of trainable parameters of DCCN. 
	The DCCN-based OFDM receiver comprises a channel equalizer followed by a basic receiver. 
	The hidden layers of DCCN are named after the signal processing modules in a legacy OFDM receiver, while they might function differently. 
	The forward network generalizes the signal processing in a legacy receiver and ensures that the search space of $\bbTheta$ contains at least an ALMMSE receiver by adding computational redundancy, which can be minimized during training by introducing a regularization loss. 
	Structural redundancy is included in the flow-graph for research, and a simplified flow-graph is proposed in Section~\ref{sec:eval:alt} for deployment.
	
	\subsection{Guidelines for Complex-valued Layers}\label{sec:ai:complex}
	
	\begin{figure}[!t]
		\vspace{-0.1in}
		\centering
		\subfloat[]{
		    \includegraphics[height=1.3in]{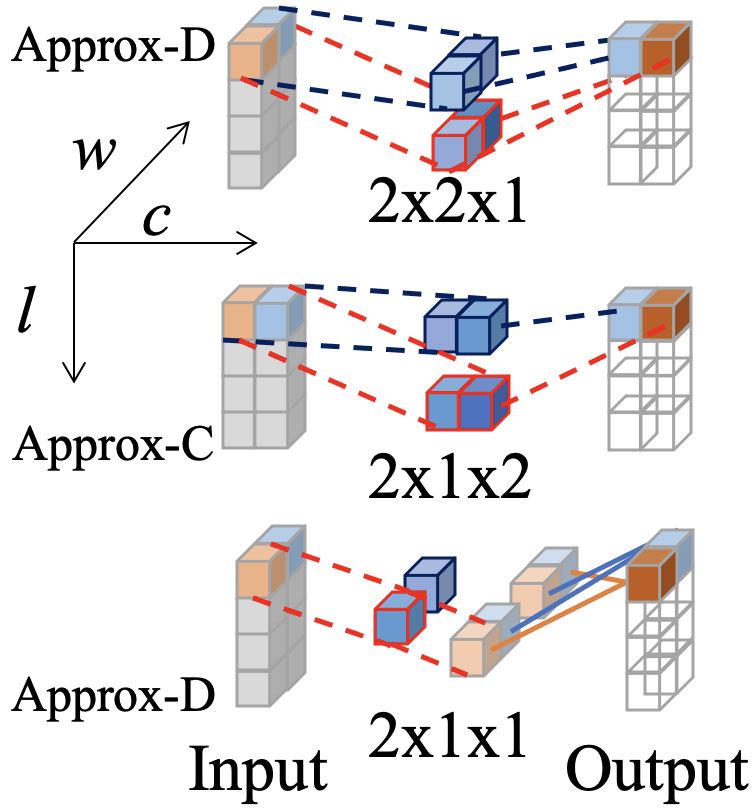}\label{fig:model:cconv0}
        }        
		\subfloat[]{
		    \includegraphics[height=1.3in]{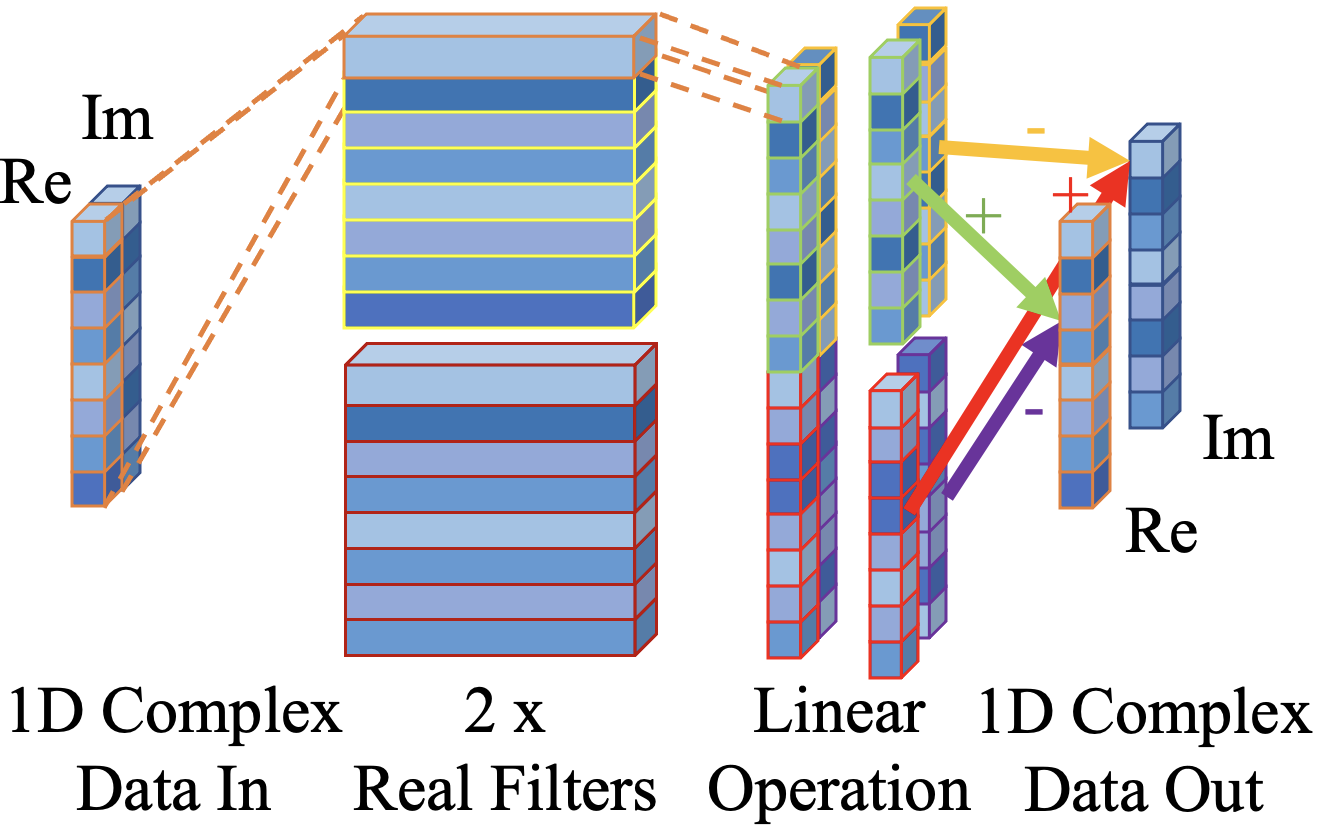}\label{fig:model:cconv1}
        }
		\caption{(a) Three implementations of 1D complex-valued Conv (C-Conv) layer ($1\times1\times1$) that preserve its expressive power, (b) Implementation of a 1D C-Conv layer ($8\times8\times1$) based on a 2D real-valued Conv layer ($16\times(1,8)\times1$)}\vspace{-0.1in}\label{fig:model:cconv}
	\end{figure}
	
	{Omitting the bias and activation, a complex-valued neuron can be expressed as \cite{hirose2012complex}:
	\begin{equation}\label{eq:cmult}
        \begin{bmatrix}
         \Re{x^{OUT}}\\ 
         \Im{x^{OUT}} 
        \end{bmatrix}
        =
        \begin{bmatrix}
         a & -b\\ 
         b & a 
        \end{bmatrix}
        \begin{bmatrix}
         \Re{x^{IN}}\\ 
         \Im{x^{IN}} 
        \end{bmatrix}\;,
	\end{equation}
	where $x^{IN},x^{OUT}\in\mathbb{C}$ and weights $a,b\in\mathbb{R}$ are the real and imaginary parts of a complex weight, respectively. \eqref{eq:cmult} can be approximated by a dense layer of $\mathbb{R}_{2\times 2}$ with approx-D input and trained weights $\bbw\approx 
	\big(\begin{smallmatrix}
      a & -b\\
      b & a
    \end{smallmatrix}\big)$. 
    Generalizing \eqref{eq:cmult} to higher dimensions, we have the following principles: 
    \begin{remark}\label{mk:dense}
	A dense layer of $\mathbb{C}_{Z_i\times Z_o}$ can always be approximated by a dense layer of $\mathbb{R}_{2Z_i\times 2Z_o}$ with full expressive power.
	\end{remark}
	\begin{remark}\label{mk:conv}
	A complex-valued convolutional (C-Conv) layer of size $f_n\times f_s\times f_c$ (stands for $f_n$ filters of shape, $f_s$, and depth of $f_c$) can be approximated by a real-valued Conv layer of size $2f_n\times (f_s,2)\times f_c$ for approx-D input or $2f_n\times f_s\times 2f_c$ for approx-C input without following \eqref{eq:cmult}, and can be exactly implemented by a real-valued Conv layer of size $2f_n\times f_s\times f_c$ with approx-D input by following \eqref{eq:cmult}. 
	\end{remark}
	}

	{By following \eqref{eq:cmult}, the exact implementation of CVNN layers reduces unnecessary degree of freedom and only needs half of the trainable real-valued weights that is required by the approximations.
	More specifically, three exemplary implementations of a 1D C-Conv layer of size $1\times1\times 1$ are given in Fig.~\ref{fig:model:cconv0} to illustrate Remark~\ref{mk:conv}. 
	All three implementations meet the expressive requirement with the same computational complexity (four multiplications and two additions per complex sample). Each of the first two approximations requires four weights while the third exact implementation only needs two weights.
	The exact implementation has better spatial complexity and training efficiency with a reduced search space for the optimizer.
	For practitioners, an approximated C-Conv layer will be less expressive than the exact one if it fails to meet the required input format or the minimal size, e.g., the shape and number of filters.
	For complex-valued activation principles, we refer readers to \cite{hirose2012complex,trabelsi2018deep}.
	} 
	
	In Fig.~\ref{fig:model:cconv1}, the implementation of our DFT-Like C-Conv layer is illustrated by an example of 1D C-Conv layer of size $8\times8\times1$ implemented by a 2D real-valued Conv layer of size $16\times(1,8)\times1$. 
	In this work, real and imaginary parts of a complex tensor are in the last dimension.

	\subsection{Basic DCCN Receiver}\label{sec:ai:single}
	
	The basic DCCN receiver is an OFDM receiver without a channel equalizer, as illustrated on the right half of Fig.~\ref{fig:flowgraph}. 
	The forward network of the basic receiver begins with an optional slicing for dropping the CP, followed by a C-Conv layer of size $N\times S\times1$ (or $N\times N\times1$ if CP is dropped), which is designed to transform time-domain OFDM symbols, $\hat{\bbx}_{cp}$, to frequency domain $\hat{\bbX}$ and exploit CP for SNR gains.
	Next, a complex dense layer $\mathbb{C}_{FN\times D}$ is designed to extract all the data REs $\hat{\bbX}_{D}$ from a coherence slot. 
	The rest of the forward network is essentially a classifier that converts IQ samples to soft bits, where an input IQ sample is treated as a vector of two real numbers, $\mathbb{C}\Rightarrow\mathbb{R}^2$. 
	The extracted IQ vector and its non-linear (Leaky ReLU) activation, $A0$, are concatenated to a tensor of shape $[B,D,4]$ and fed to a small dense layer of $\mathbb{R}_{4\times 2^m}$ followed by another Leaky ReLU activation, $A1$, of which the output tensor is reshaped to $[B,D,m,2]$ and then activated by a softmax function along its last dimension to produce a soft bit--a vector of likelihoods of $\pm1$. 
	$B$ is the number of slots in a batch of input signal. Since the channel coding is out of our scope, the output bits are obtained by hard decisions on soft bits. 
	In Section~\ref{sec:eval:alt}, variations of the forward network are tested.
	
	\begin{figure}[!t]
		\vspace{-0.1in}
		\centering
		\includegraphics[width=\linewidth]{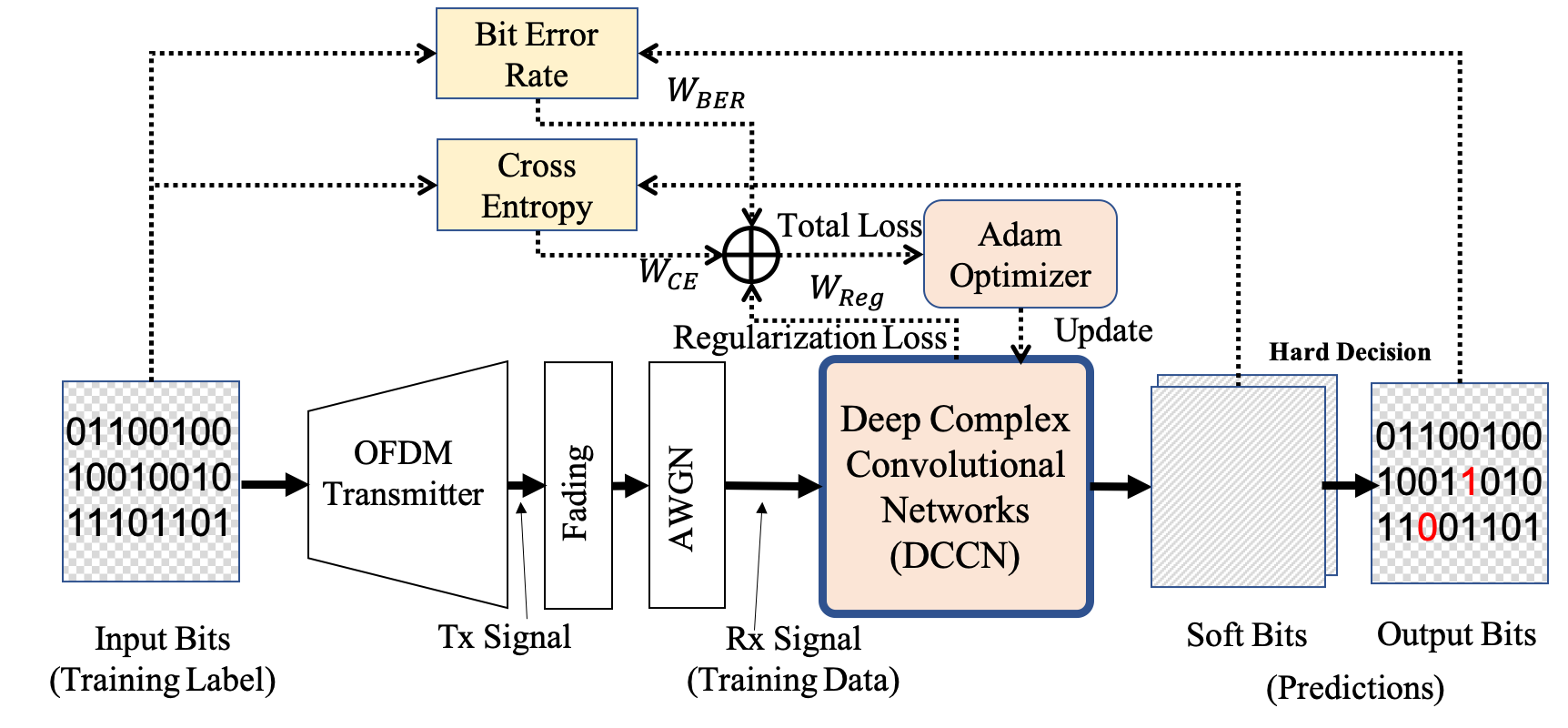}
		\vspace{-0.1in}
		\caption{Block diagram of the DCCN training System.}\label{fig:model:train}
		\vspace{-0.1in}
	\end{figure}

	\subsection{DCCN Channel Equalizer}\label{sec:ai:multi}
	The DCCN equalizer, with the input of $\bby_{cp}$ and the output of $\hat{\bbx}_{cp}$, is prepended to the basic DCCN receiver, as illustrated on the left half of Fig.~\ref{fig:flowgraph}. 
	The forward network of the DCCN equalizer contains four submodules: 
	The first submodule comprises a dense layer of $\mathbb{C}_{S\times N}$ (or $\mathbb{C}_{N\times N}$ if CP is dropped), followed by a C-Conv layer of $N\times N\times 1$, and it converts $\bby_{cp}$ to the frequency domain $\bbY$.
	The second submodule estimates the channel frequency response $\hat{\bbH}$ with four dense layers followed by a 2D complex filter. 
	The third submodule performs equalization with an element-wise complex division $\hat{\bbX}=\bbY/\hat{\bbH}$. 
	Finally, $\hat{\bbX}$ is converted to  $\hat{\bbx}$ by an IDFT-Like 1D C-Conv layer  of $N\times N\times 1$, and $\hat{\bbx}_{cp}$ is recovered by adding back CP with a dense layer of $\mathbb{C}_{N\times S}$.

	In the channel estimation submodule, the first dense layer of $\mathbb{C}_{FN\times P}$ is designed to locate pilots and estimate channel coefficients on pilots $\hat{\bbH}^{P}_{LS}$.
	Then, $\hat{\bbH}$ is obtained by an interpolation of $\hat{\bbH}^{P}_{LS}$ to the entire coherence slot and channel estimation in the next three dense layers and a 2D filter of size $(F,N)$, which resembles the LRA-LMMSE \cite{edfors1998ofdm}:
	\begin{equation}\label{eq:lralmmse}
		\hat{\bbH}_{LRA}=\bbU\bbD_p\bbU^{H}\hat{\bbH}_{LS}\;,
	\end{equation}
	where $\bbD_p$ is a diagonal matrix with entries $\delta_{k}=\frac{\lambda_{k}}{\lambda_{k}+\beta/\alpha}$ for $k \in [1,p]$, and $\delta_{k}=0$ for $k\in[p+1,N]$, $\bbU$ is a unitary matrix containing the singular vectors of $\bbR_{\bbH\bbH}$. 
	Instead of setting $\bblambda$, $p$, $\alpha$, and $\bbU$ explicitly \cite{hung10pdp,savaux2015joint,Zhou09lmmse}, the LMMSE matrices in \eqref{eq:lralmmse} are trainable parameters of multiple dense layers to be learned from data. 
	Note that the dimension of these dense layers are of a coherence slot for ISI mitigation. 
	In Section~\ref{sec:eval:fading}, our DCCN receiver also exhibits an error floor like ALMMSE with a prescribed SNR \cite{edfors1998ofdm,savaux2015joint,Zhou09lmmse}. 
	Other numbers of the dense layers of $\mathbb{C}_{FN\times FN}$ are tested in Section~\ref{sec:eval:alt}.

	\begin{figure}[!t]
		\centering
		\includegraphics[width=0.95\linewidth]{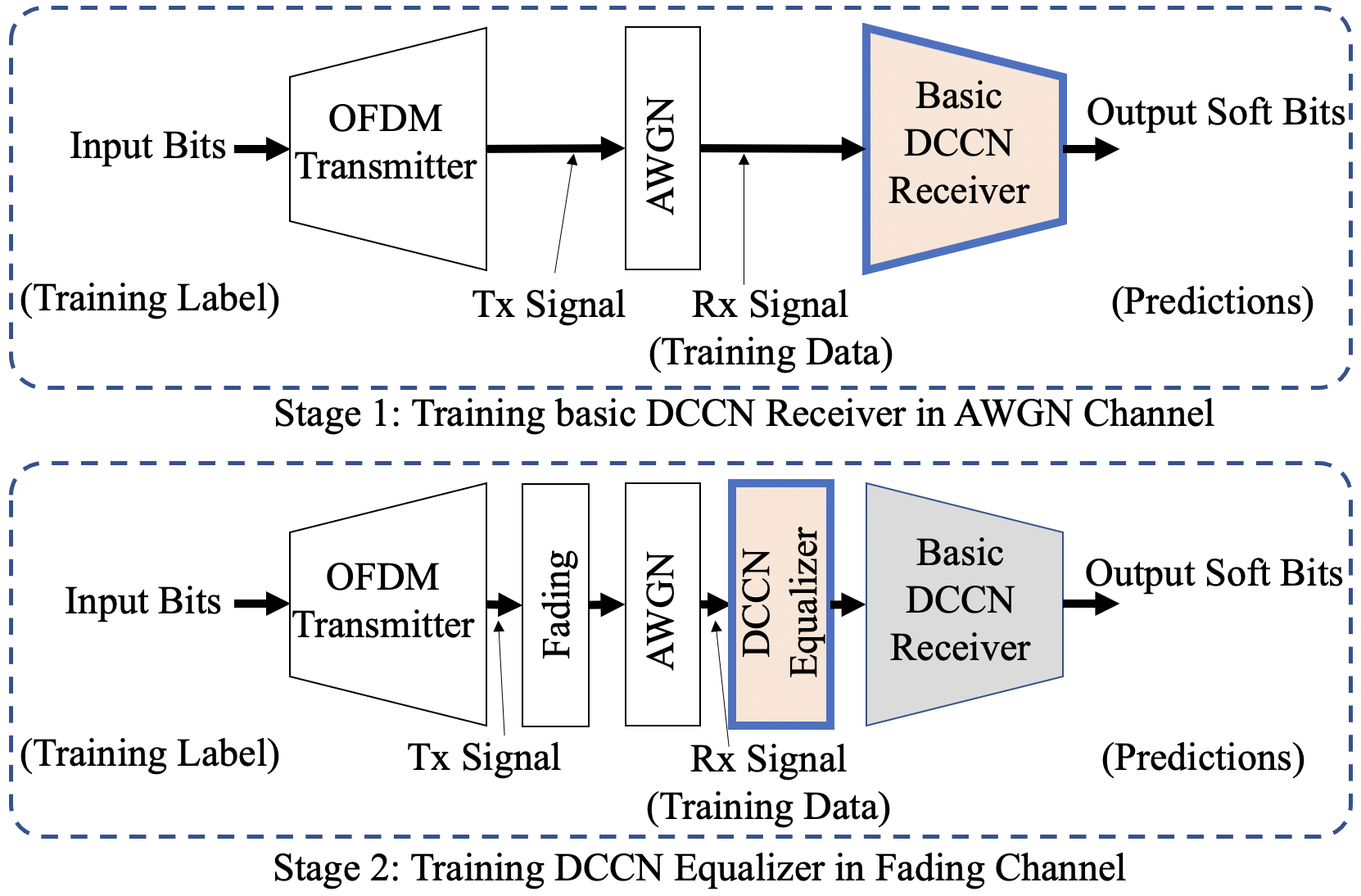}
		\caption{Training the basic receiver and the equalizer in two stages.}\label{fig:learn:trans}
		\vspace{-0.1in}
	\end{figure}
	
	
	{Considering the number of OFDM symbols per coherence slot $F$ as a constant in a protocol, the asymptotic computational complexity of the DCCN receiver} is $\ccalO(N^2)$ (or $\ccalO(NS)$ for DCCN with CP), since DCCN is composed of cascading layers without any loops. 
	The 1D complex convolutional layer and the fully connected layer have a maximum complexity of $\ccalO(N^2)$ (or $\ccalO(NS)$ for with CP). 
	{If $F$ is considered as a variable, the complexity becomes $\ccalO(N^2F^2)$.}
	
	\subsection{2-Stage Training}\label{sec:cdnn:train}

	The training setup of the DCCN receiver is illustrated in Fig.~\ref{fig:model:train}. An online random generator creates a random binary stream $\bbb$, which is translated by an OFDM transmitter into time domain OFDM symbols as the transmit signal, $\bbx_{cp}$. The received signal $\bby_{cp}$ is created by a channel model that adds fading and noise to $\bbx$. $\bby_{cp}$ and $\bbb$ are the training data and labels, respectively. The outputs of DCCN model are the soft bits, $\tilde{\bbb}$, and the output bits, $\hat{\bbb}\in\{\pm1\}$, which are generated by hard decision on $\tilde{\bbb}$. 
	{The loss function $\ccalL(\bbb,\tilde{\bbb},\hat{\bbb},\bbTheta)$ is the weighted sum of the cross entropy (CE) loss and regularization loss $\ccalL_{reg}(\bbTheta)$:
	\begin{equation}\label{eq:loss}
		\ccalL(\bbb,\tilde{\bbb},\hat{\bbb},\bbTheta) = \ccalL_{CE}(\bbb,\tilde{\bbb}) +
		\epsilon\ccalL_{reg}(\bbTheta)\;,
	\end{equation}
	where $\epsilon\ll 1$ is a small constant.}
	The cross entropy loss, $\ccalL_{CE}(\bbb,\tilde{\bbb})$, is the average cross entropy of the training labels $\bbb$, and the soft bits $\tilde{\bbb}$. 
	During  training, $\bbTheta$ of DCCN receiver is randomly initialized and updated by an Adam optimizer that is running back-propagation based on the loss function in \eqref{eq:loss}.
	
	
	
	It is difficult and may be impractical to train the DCCN receiver directly in multipath fading channels due to the severe distortion shown in Fig.~\ref{fig:channel:EPA}. Therefore, a transfer learning scheme, as illustrated in Fig.~\ref{fig:learn:trans}, is developed to train the basic receiver and the equalizer in two stages. 
	In stage 1, the basic receiver is trained in an AWGN channel only. 
	In stage 2, the flow-graph of the DCCN equalizer is first prepended to the pre-trained basic receiver in a TensorFlow session dedicated for graph-editing. Then, the equalized DCCN receiver is loaded and trained in another session, in which the trainable parameters of the basic receiver is frozen and the channel fading is included to generate the training data. The loss function is the same throughout stages 1 and 2. The graph-editing technique enables back-propagation when the second half of the forward network is frozen. 
	{Note that the two-stage training approach can increase the data efficiency by reusing the same pre-trained basic receiver in stage 2 for different fading settings. Similarly, the trained model from stage 2 can be fine-tuned for different realistic channels.}
	
	
	To improve training efficiency, several techniques are employed. The training data is fed to the model in mini batches, which leverages the high throughput of the parallel processing in the graphics processing units (GPUs) and minimizes the latency in memory copying. In the programming of NumPy-based OFDM transmitter and fading modules, data processing is vectorized and large loops are avoided. The learning rate is decayed exponentially for fine-tuning as the training proceeds. The random training labels and corresponding training data are generated online rather than feeding a pre-generated training dataset repeatedly. Therefore, we use an \textit{iteration} instead of an epoch (a full training pass over the entire dataset) to describe the outermost loop in training. An early stop mechanism is employed on top of a maximum number of training iterations to end the training if the key performance metric (i.e., BER) has not improved after a fixed number of iterations. 
	
	\subsection{Channel Settings in Training}
	A clear guidance does not exist for setting the training SNR for a DL-based PHY. 
	{We use different SNR configurations in the two stages. 
	In stage 1, the SNR does not influence the optimal basic receiver but the effectiveness of training. The noise in the channel creates bit errors that drive the gradient descent and acts as a regularizer to prevent over-fitting.
	A higher SNR requires a larger size of the mini-batch to generate the same amount of bit errors, which could lower the data efficiency and prolong the convergence of training. 
	On the other hand, low SNRs may hide small demodulation bias in relatively large BERs, i.e., consistently more $1$s than $-1$s in the output bits. To this end, we recommend a training SNR value of {$E_b/N_o=5$}{~dB} \cite{OShea16a} for the training stage 1, specifically, the SNR {$\eta_{i}=5m$}{~dB} is selected for $m$-ary modulation.}
	
	In the training stage 2, the optimal DCCN equalizer depends on the channel statistics such as SNR and PDP. 
	For better generalizability of the trained model, a mixture of SNRs and fading models is employed. 
	The SNR of each OFDM coherence slot is randomly selected from the working SNR regime based on a probability function that prefers high SNR values without excluding middle and low SNR values. 
	For example, $\eta_{i}\in\{0, 3, \dots, 30\}$ with $P(\eta_{i}\geq17)=90\%$. 
	The rationale is that the DCCN equalizer resembles a fast ALMMSE algorithm \cite{Zhou09lmmse}, which has preferable performance when designed for high SNR (i.e., 20~dB) other than low SNR (i.e., 5~dB). 
	Meanwhile, since LMMSE is robust to PDP mismatch, the optimal points of the DCCN equalizer for different PDPs would be located closely. 
	A mixture of Rayleigh fading models improves not only the generalizability but also the convergence time.  
	The fading models with shorter delay spread can smooth the overall loss landscape to help the optimizer overcome the local minima associated with those with richer multipath. 
	In Sections~\ref{sec:eval:gen} and~\ref{sec:eval:alt}, models trained with different channel settings in both stages are compared.
	
	\section{Evaluation Results} \label{sec:eval}
	
	\subsection{Methodology}
	The DCCN receiver is compared to  legacy receivers with different channel estimators \cite{dlofdm18} in numerical evaluations.
	We first present the results of the DCCN basic receiver for $m$-ary Quadrature Amplitude Modulation (QAM) modulation ($m\leq 4$) in AWGN channels, then the equalized DCCN receiver in Rayleigh fading channels with different settings of PDP, ISI leakage, and mobility. 
	We use the notations DCCN and DCCN-CP to refer to DCCN receivers without and with CP exploitation, respectively.

	
	\begin{table}[!t]
		\renewcommand{\arraystretch}{1}
		\caption{Configurations of The Evaluated OFDM System} 
		\vspace{-0.1in}
		\label{tab:ofdm}
		\centering
		\footnotesize
		\begin{tabular}{p{0.29\linewidth}|p{0.56\linewidth}}
			\hline
			$FFT$ Size  & $N=64$   \\ \hline
			Sample rate (Msps) & 0.96   \\ \hline
			Guard SCs per symbol & $14$ (Outer) + $2$ (DC)   \\ \hline			
			Pilot REs per slot & $P=16$ \\ \hline
			Data REs per slot & $D=320$ \\ \hline
			Subcarrier bandwidth & {$15$~KHz}  \\ \hline
			Symbols per slot & {$F=7$ }  \\ \hline
			CP length & {long: $0.25N$, short: $0.07N$ } \\ \hline
			PAPR limit & {$9$~dB} \\ \hline
			Pilot & {LTE downlink frame pilot pattern \cite{3gpp36211} with constant value $\sqrt{1/2}(1+i)$ } \\ \hline
			Modulation & {QAM with orders of $2,4,8,16$, Gray code}  \\ \hline
			Rayleigh fading model & {Flat, EPA, EVA, ETU \cite{3gpp36104}}  \\ \hline
			Channel coherence & {$T_c>1ms$, $B_c\geq200KHz$ \cite{3gpp36104}}  \\ \hline
			{Max Doppler spread} & {$F_d=0$~Hz by default, or if specified {EVA $70$~Hz, ETU $70$~Hz, $300$~Hz} \cite{3gpp36104} } \\ \hline
		\end{tabular}
	\end{table}
	
	The evaluated OFDM system emulates a simplified LTE downlink frame structure \cite{3gpp36211} as detailed in Table \ref{tab:ofdm}, in which the DFT size is $64$, the total number of guard subcarriers (SCs) at the edge and direct current (DC) is $16$, the sampling rate is $0.96$~MHz, and the bandwidth of a subcarrier is $15$~KHz. 
	A coherence slot contains 7 OFDM symbols, in which the numbers of REs assigned to pilot and data are $P=16$ and $D=320$, respectively. 
	{A scattered pilot pattern, which is consistent with LTE protocol as illustrated in the rightmost of Fig.~\ref{fig:ofdm:pilot}, is used, and the pilot signal has a constant value of $\sqrt{1/2}(1+i)$.} 
	QAM with orders of 2, 4, 8, 16 and Gray code are used for constellation mapping, where the maximum amplitude of the constellation is $1$.
	The peak to average power ratio (PAPR) of the OFDM waveform after the transmitter is limited to $9$~dB. 

	\begin{table}[!t]
		\renewcommand{\arraystretch}{1}
		\caption{Training Configurations for $m$-ary Modulation} 
		\vspace{-0.1in}
		\label{tab:train}
		\centering
		\footnotesize
		\begin{tabular}{l|c|c}
			Setting & Basic Receiver & DCCN Equalizer \\ \hline
			Maximum iterations &  $1200m$ & $4000m$ \\ \hline
			Early stop window & 200 iterations & 200 iterations \\ \hline
			Initial learning rate & 0.001 & 0.001 \\ \hline
			Learning rate decay & \multicolumn{2}{c}{Exponential, rate 2\%, step 500 (mini-batches) } \\ \hline
			SNR (dB) & {$5m$}  & Customized random variable\\ \hline
			Iteration & \multicolumn{2}{c}{$200$ mini-batches}  \\ \hline
			Mini-batch size &  \multicolumn{2}{c}{$72 \times 320 \times m$ bits}  \\ \hline
			Testing bits per SNR & \multicolumn{2}{c}{$20000 \times 320 \times m$ bits }  \\ \hline			
			Testing SNR &  \multicolumn{2}{c}{-10 to 30~dB} \\ \hline
			Optimizer & \multicolumn{2}{c}{SGD with Adam} \\ \hline
		\end{tabular}
	\end{table}
	
	The training configurations for the basic receiver and equalizer with $m$-ary modulation are listed in Table~\ref{tab:train}. 
	A stochastic gradient descent-based Adam optimizer with a mini-batch size of 72 coherence slots are used. 
	In each iteration, a new random bit stream (training labels) is generated and converted to the training data by a legacy OFDM transmitter and a channel model. 
	Since training labels are random, the loss function for early stop mechanism is based on the training data rather than a separate test data set. 
	Training terminates either by reaching the maximum number of iterations, or triggering an early stop which mostly happens in practice. 
	The learning rate is set to $0.001$ initially and decayed by 2\% every $500$ mini-batches (steps) or $2.5$ outer iterations. 
	The SNR is set according to the description in Section~\ref{sec:cdnn:train}. 
	Each outer training iteration contains $200$ mini-batches, and the testing data is of $20,000$ coherence slots for each SNR point from $-10$ to $30$~dB. 
	{The Rayleigh fading for training stage 2 is set as alternating models of flat fading, EPA, EVA, and ETU in consecutive slots.}
	
	\begin{figure}[!t]
		\centering
		\vspace{-0.2in} 
		\subfloat[Stage 1: Receiver]{	
			\includegraphics[width=0.48\linewidth]{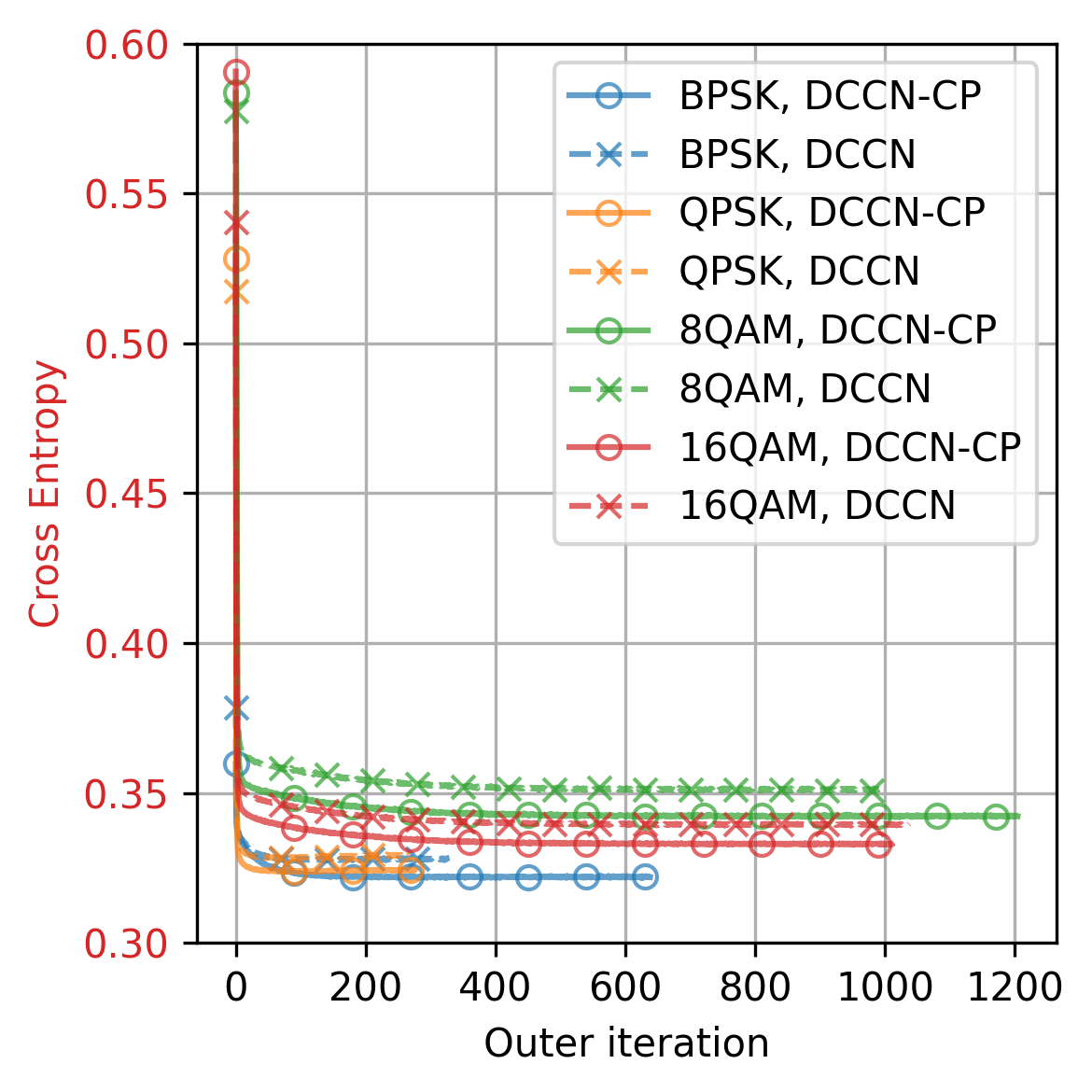}\label{fig:loss:awgn}
		}
		\subfloat[Stage 2: Equalizer]{	
			\includegraphics[width=0.48\linewidth]{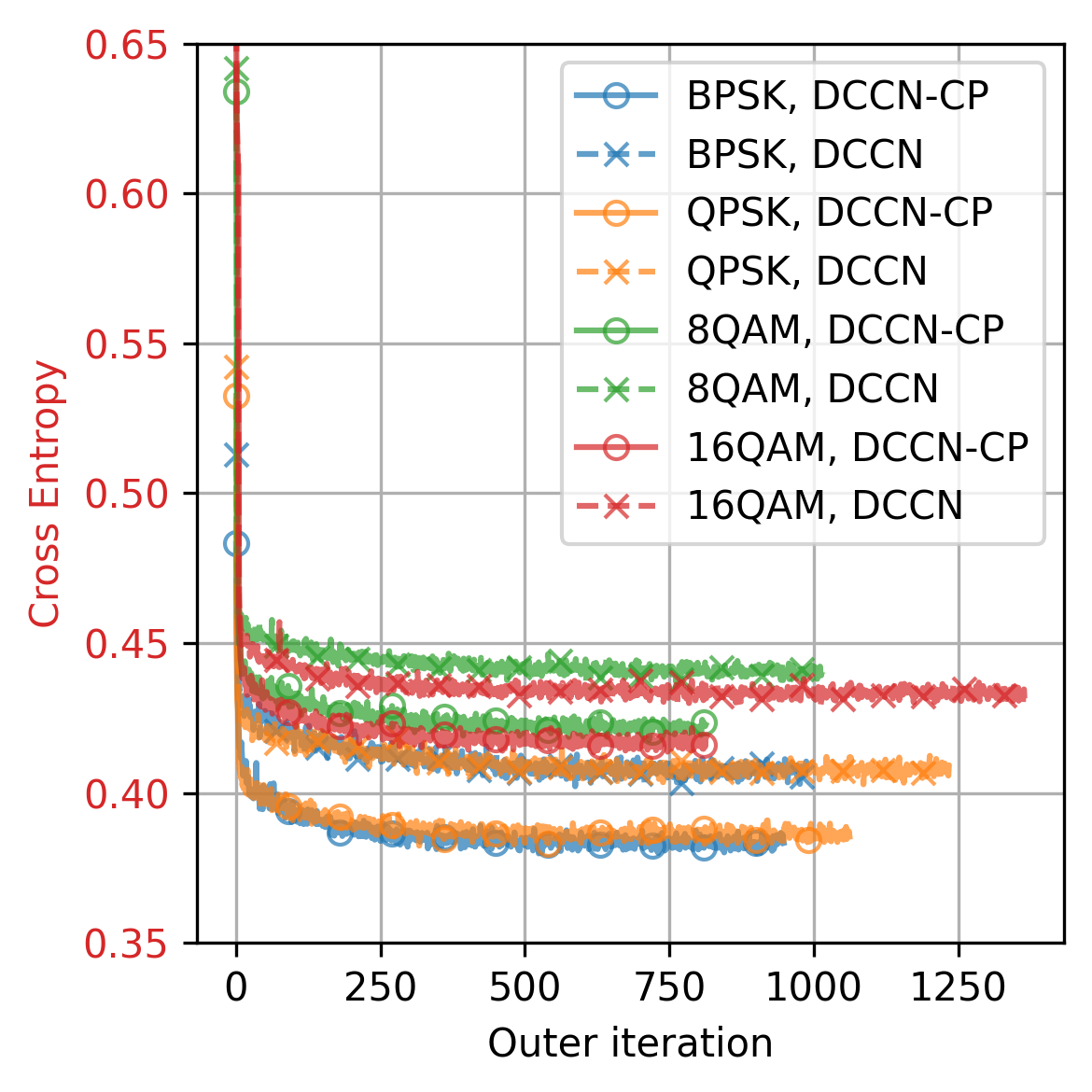}\label{fig:loss:flat}
		}
		\caption{Evolution of cross entropy loss in the training of DCCN models}
		\label{fig:loss}
		\vspace{-0.1in}
	\end{figure}

	Training a DCCN model lasts for  $250$--$1300$ iterations, as shown in Figs.~\ref{fig:loss}. 
	The training process starts with a quick fitting followed by a long fine-tuning phase: the cross entropy first decreases drastically in the first 10--50 iterations, then decreases slowly but steadily until hitting a floor. 
	The source code is available at \cite{dlofdm18}.
	
	\subsection{Additive White Gaussian Noise Channel}\label{sec:eval:model}
	
\begin{figure}[!t]
		\centering
		\includegraphics[width=\linewidth]{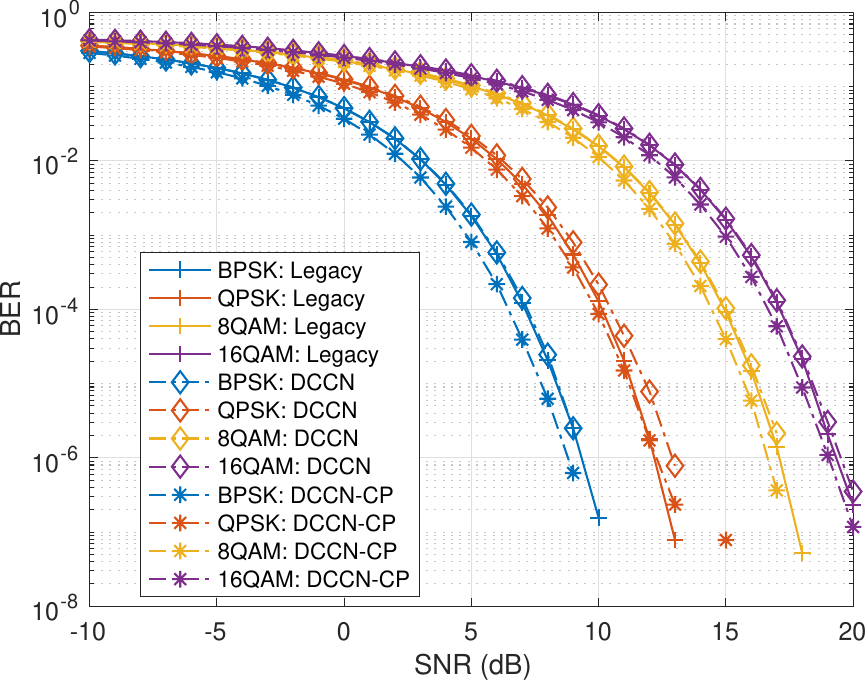}
		\vspace{-0.1in} 
		\caption{BER of legacy and DCCN OFDM receivers vs SNR, in AWGN channel, with long CP.}
		\vspace{-0.1in}
		\label{fig:ber:awgn}
	\end{figure}

	The BERs of our basic receivers and the legacy OFDM receiver \cite{dlofdm18} with modulations of BPSK, QPSK, 8QAM and 16QAM in AWGN channels with SNR from $-10$ to $20$~dB are presented in Fig.~\ref{fig:ber:awgn}, where long CP is considered for DCCN-CP.
	{Without the help of CP, we do not expect DCCN to outperform the legacy demodulation since the latter is optimal in an AWGN channel.}
	For BPSK, 8QAM and 16QAM, DCCN performs closely to the legacy receiver with a negligible gap ($\leq 0.16$~dB) when BER is below $10^{-5}$. 
	However, for QPSK, DCCN begins to underperform the legacy receiver at the BER level of $10^{-2}$ and the gap is increased to $0.7$~dB at BER of $10^{-6}$. 
	On the other hand, the DCCN-CP consistently outperforms DCCN by a gap that decreases from $0.7$~dB for BPSK to $0.5$~dB for 16QAM.
	For QPSK, DCCN-CP starts to underperform legacy receiver at BER of $10^{-6}$. 
	Note that when BER is very small, e.g., $\leq 10^{-5}$, the relative error due to limited data size in simulation increases.
	The CP carries information on all subcarriers and experiences independent random noise, thus can theoretically improve the power of data signal by ${N_{cp}D}/(FN^2)$, which corresponds to an increase of $0.75$~dB on SNR in AWGN channel for a CP length of $0.25N$. 
	DCCN-CP brings an improvement of SNR to DCCN very close to $0.75$~dB for BPSK.
	It shows that using C-Conv layer to transform time-domain OFDM symbol to frequency-domain has the advantage of exploiting redundancy carried by CP, which can not be accomplished by explicit DFT.

	\subsection{Rayleigh Fading Channels}\label{sec:eval:fading}
	
	Next, the equalized DCCN receivers are evaluated in Rayleigh fading channels. 
	The benchmarks are legacy OFDM receivers with different channel estimators: LS estimator (LS-Spline), LS estimator enhanced with CP based on analytical approach in \cite{Quadeer10} (LS-CP), the ideal LMMSE, and an ALMMSE. 
	{The channel covariance matrix is updated per OFDM symbol for the ideal LMMSE as $\bbR_{\bbH\bbH}=\bbH_{*,i}^{}\bbH_{*,i}^H$ based on true channel realization $\bbH_{*,i}$, and per slot for ALMMSE as $\hat{\bbR}_{\bbH\bbH}=\frac{1}{F}\hat{\bbH}_{LS}^{}\hat{\bbH}_{LS}^H$.
	The $\hat{\bbH}_{LS}$ in the ideal LMMSE and ALMMSE is from LS-Spline.}
	The tested ALMMSE represents the upper bound of ALMMSE of complexity $\ccalO(N^2)$ \cite{Zhou09lmmse} with full rank approximation and perfect SNR estimates. 
	The ISI is set to be only between consecutive OFDM symbols in the same coherence slot. 
	For simplicity, BPSK modulation and perfect synchronization are considered at the receiver. 
	{An independent channel realization is generated per coherence slot and is invariant within the slot by default ($F_d=0$~Hz) \cite{HYe18}. If $F_d>0$~Hz, the time-varying channel is generated by the Jake's model and the fading technique of sum of $48$ sinusoids \cite{matlabfading}}.
	The flat fading and 3GPP multipath fading models \cite{3gpp36104} are evaluated.
	With the DFT size of $64$, the channel filter lengths, $L$, for flat fading, extended pedestrian A model (EPA), extended vehicular A model (EVA), and extended typical urban model (ETU) are $1$, $9$, $11$ and $13$, respectively.

	\begin{figure}[!t]
		\centering
		\subfloat[]{	
			\includegraphics[width=\linewidth]{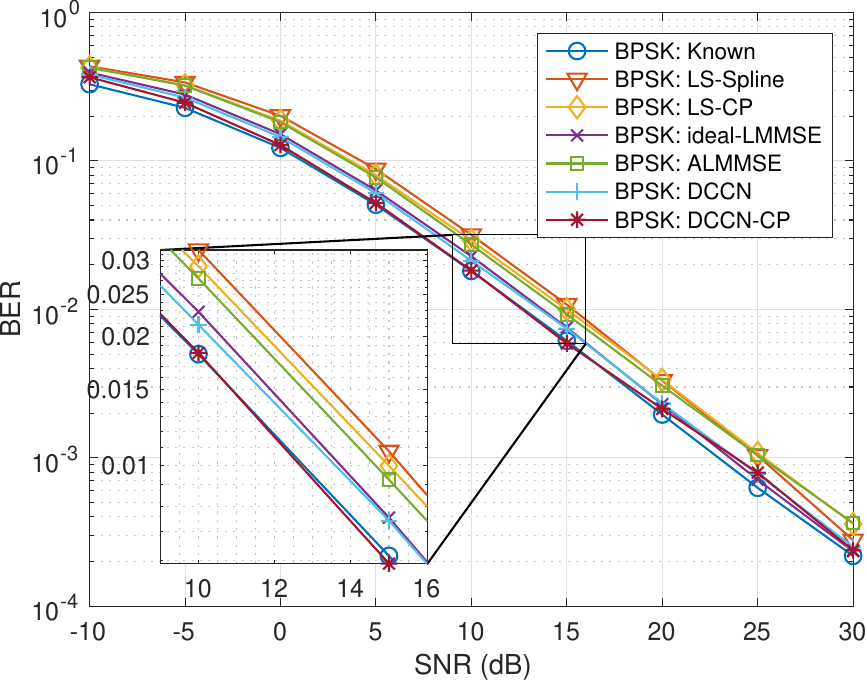}\label{fig:ber:fad:flat}
		}\\ \vspace{-0.1in} 
		\subfloat[]{	
			\includegraphics[width=\linewidth]{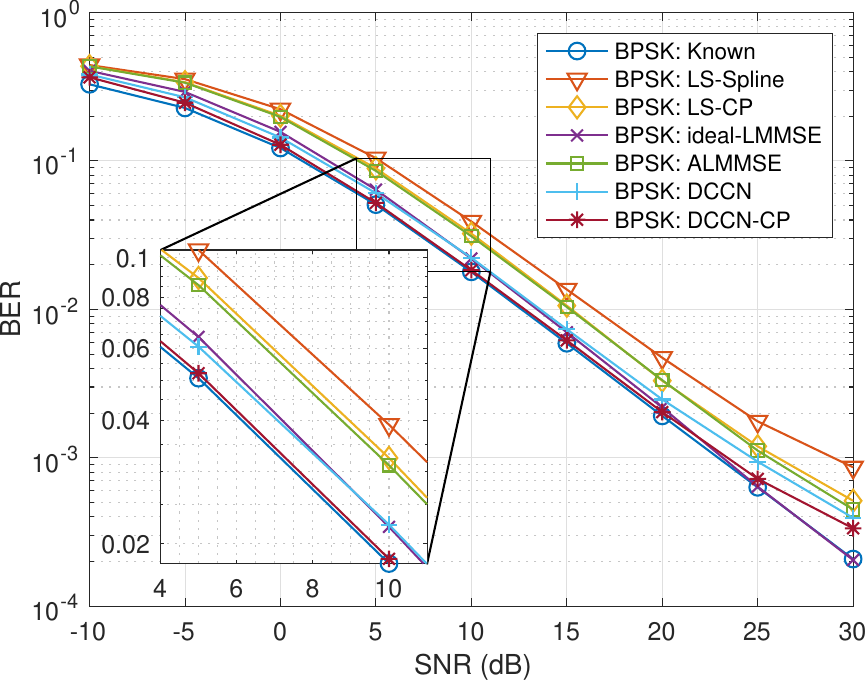}\label{fig:ber:fad:epa}
		}
		\caption{BER of equalized DCCN receivers and benchmarks with long CP $N_{cp}=16$ in low-mobility Rayleigh fading channels: (a) flat fading  $L=1$, and (b) multipath fading with low RMS delay spread (EPA \cite{3gpp36104}), $L=9$.}\vspace{-0.1in}
		\label{fig:ber:fad}
	\end{figure}
	
	\begin{figure}[!t]
		\centering
		\subfloat[]{	
			\includegraphics[width=\linewidth]{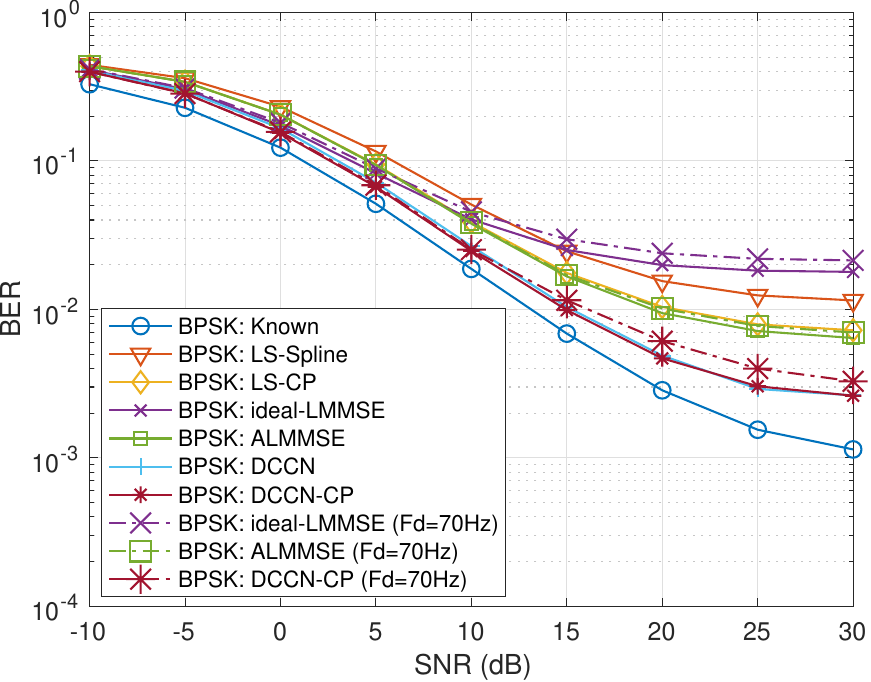}\label{fig:ber:eva:long}
		}\\ \vspace{-0.1in} 
		\subfloat[]{	
			\includegraphics[width=\linewidth]{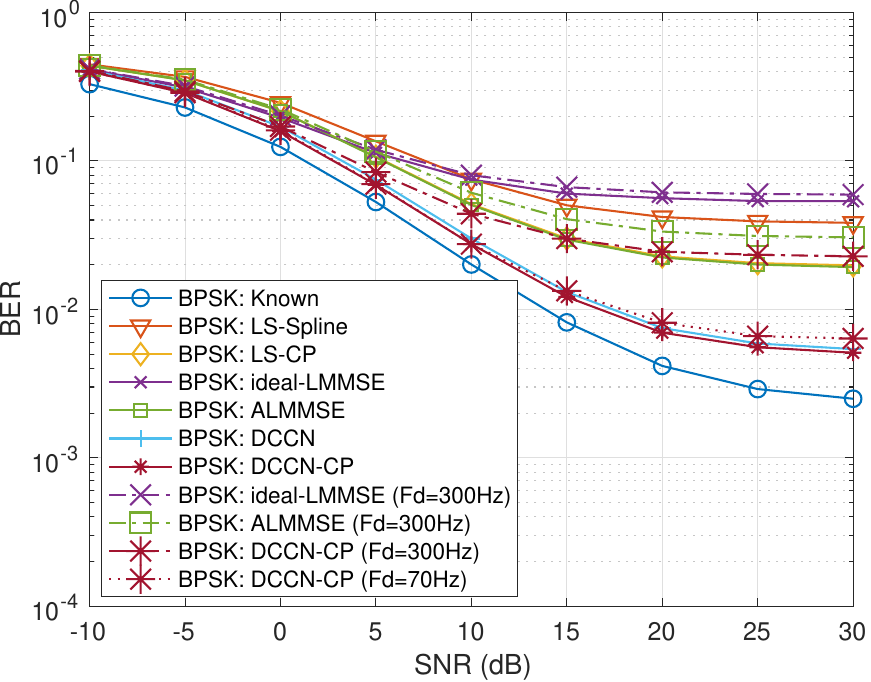}\label{fig:ber:eva:short}
		}
		\caption{BER of equalized DCCN receivers and benchmarks with short CP $N_{cp}=4$ in multipath Rayleigh fading channels with large RMS delay spread and leakage of ISI: (a) EVA \cite{3gpp36104}  $L=11$, and (b) ETU \cite{3gpp36104} $L=13$.}\vspace{-0.1in}
		\label{fig:ber:eva}
	\end{figure}

\begin{figure}[!t]
		\centering
		\includegraphics[width=0.99\linewidth]{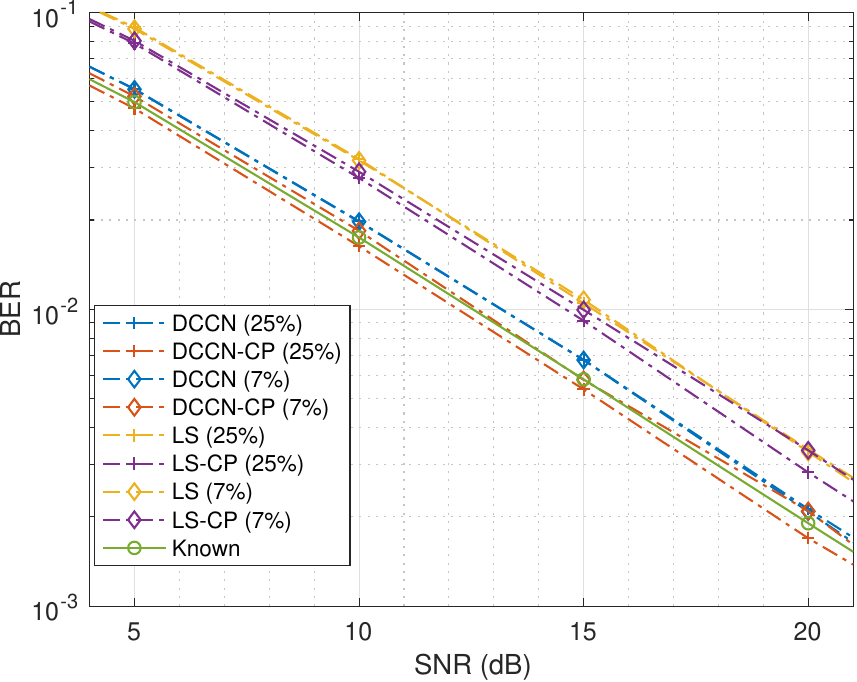}
		\vspace{-0.1in} 
		\caption{BER of DCCN receivers with different CP lengths vs SNR in flat fading channel. The DCCN equalizer is trained only in flat fading channel. The gains of DCCN-CP receiver with long CP ($25\%N$) and short CP ($7\%N$) are $0.88$~dB and $0.33$~dB, respectively. The DCCN receiver that drops CP is on average $0.5$~dB worse than the receiver with known channel information. Legacy receiver with CP-enhanced LS equalizer \cite{Quadeer10} gains $0.7$~dB and $0.3$~dB over baseline LS equalizer in long and short CPs, respectively.}\vspace{-0.1in}
		\label{fig:ber:cplength}
	\end{figure}

	In Rayleigh fading channels of small delay spread, such as flat fading and EPA model \cite{3gpp36104}, the BER performances of the equalized DCCN receivers and benchmarks with long CP are presented in Figs.~\ref{fig:ber:fad:flat} and~\ref{fig:ber:fad:epa}, respectively. 
	Since the length of channel filter $L$ is smaller than length of CP $16$, the ISI can be completely removed by dropping CP. 
	Compared to the ideal LMMSE, the DCCN receiver that drops CP is similar or slightly better in low to mid-range SNR regime ($\leq15$~dB), and underperforms in high SNR regime with a tiny performance gap in flat fading and a larger gap of $1$ to $3$~dB in EPA channel. 
	The DCCN-CP receiver outperforms the ideal LMMSE by around $1$~dB and almost overlap with the ideal receiver with perfect channel estimates in the low to mid-range SNR regime, but its performance also deteriorates in high SNR regime. 
	It shows that DCCN-CP receiver is able to exploit the redundancy information carried in CP for performance gain. 
	Despite that DCCN-CP receiver coincidentally overlaps with perfect channel estimates here, its performance gain in fact depends on the amount of redundancy in CP, which is discussed later in Fig.~\ref{fig:ber:cplength}. 
	Compared to flat fading, the delay spread in EPA channel enlarges the performance gap between the ideal LMMSE and ALMMSE from $0.8$ to $1.4$~dB, and the gap between ALMMSE and LS-Spline from $0.7$ to $1$~dB, while LS-CP stays closely to ALMMSE. 
	Although LS-CP underperforms ALMMSE by about $0.2$~dB, LS-CP requires no prior information while ALMMSE is based on perfect SNR. Notice that SNR of $30$~dB may not be common in practice, and the uncertainty of tested BER at the level of $10^{-4}$ is also higher. 
	The fact that DCCN performs similar to the ideal LMMSE with prior knowledge of channel covariance matrix when ISI is completely removed, shows that DCCN can learn the channel statistics from data.
	
	In a multipath channel of large delay spread, such as EVA and ETU models \cite{3gpp36104}, the BER performance by SNR of the equalized DCCN receivers and benchmarks with short CPs are presented in Figs. \ref{fig:ber:eva:long} and \ref{fig:ber:eva:short}, respectively. 
	Notably, in the case of the length of the channel exceeds the length of CP, the ISI can not be removed by dropping CP, resulting performance degradation in high SNR regime even with perfect channel knowledge. 
	With the presence of ISI, the ideal-LMMSE underperforms LS and all the other channel estimators in mid-to-high SNR regime, similar results was found in \cite{Khlifi2011}.
	Meanwhile, by alleviating the notch in frequency in frequency-selective fading \cite{Quadeer10}, the performance gain of CP-enhanced LS estimator over the baseline LS estimator increases by the delay spread of the channels. 
	In a frequency-selective channel with large delay spread, DCCN receiver outperforms the ALMMSE, LS, and CP-enhanced LS, with a large margin in high SNR regime, e.g., leading ALMMSE by $5$~dB at BER of 0.01. 
	The DCCN equalizer can take advantage of processing a whole coherence slot instead of a single OFDM symbol, thus further reduce the ISI. 
	Similar mechanism is employed in conventional CP-enhanced equalizer \cite{Quadeer10}, in which decoded data of the previous OFDM symbol is used to mitigate the ISI in the next one. 
	DCCN can learn to mitigate ISI and frequency nulling together in frequency-selective fading channels, yielding superior performance from synergy. 
	In comparison, CP-enhancement approach in \cite{Quadeer10} can only enhance baseline LS estimation but fails to improve ideal-LMMSE and ALMMSE in our experiment. 
	
	{The BERs of the tested receivers in doubly selective channels, simulated by including Doppler spread in the EVA and ETU channels, are shown with the dashed lines in Figs.~\ref{fig:ber:eva:long} and~\ref{fig:ber:eva:short}. 
    In the mobile channels, the Doppler effect increases the temporal channel variation and causes frequency dispersion that leads to inter-carrier interference at the OFDM receiver.
	With a maximum Doppler frequency of $70$~Hz in both EVA and ETU channels, the BERs of ideal-LMMSE, ALMMSE, and DCCN-CP rise by 0.003, 0.0008, 0.001, respectively, in middle-to-high SNRs. 
	The DCCN-CP still leads by a gap similar to the static cases.
	However, with a maximum Doppler spread of $300$Hz in the ETU channel, the BERs of ideal-LMMSE, ALMMSE and DCCN-CP rise by $0.007$, $0.01$, and $0.018$, respectively, where DCCN-CP suffers the most.
	These results show that DCCN-CP is robust to channel mobility, but with a relatively low pilot density of $4.7\%$,  its advantage over legacy receivers can be compromised in high mobility scenarios.}
	
	\begin{figure}[!ht]
		\centering
		\includegraphics[width=\linewidth]{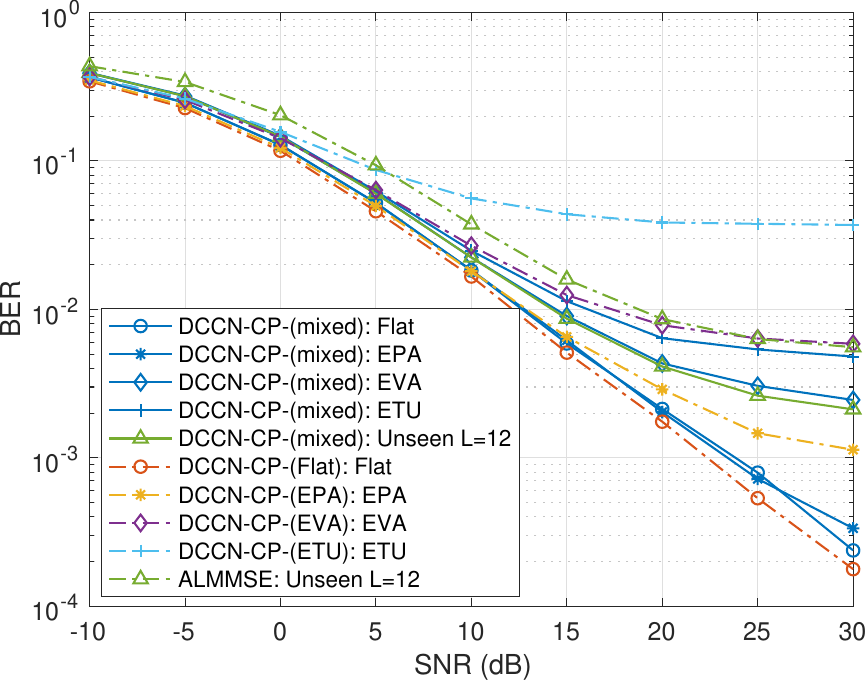}
		\vspace{-0.1in} 
		\caption{Comparison of DCCN equalizers trained with mixed Rayleigh fading models vs trained with single channel model and tested on the same channel model, the four Rayleigh fading models are: flat fading, EPA, EVA, ETU \cite{3gpp36104}}\vspace{-0.1in}
		\label{fig:ber:gen}
	\end{figure}

	The effectiveness of CP enhancement of both equalized DCCN receivers and CP-enhanced LS receivers \cite{Quadeer10} are evaluated in OFDM systems with long and short CPs, as illustrated in Fig.~\ref{fig:ber:cplength}. 
	For a fair comparison, flat fading channel is selected since no ISI will be leaked into the main OFDM symbol with short CP. The equalized DCCN receivers are trained in flag fading channels only. 
	Both the baseline DCCN and LS receivers that drop CP has identical performance regardless of the CP length. 
	In the settings of long and short CP, the DCCN-CP receiver gains $0.88$~dB and $0.33$~dB, respectively, and the CP-enhanced legacy LS receiver \cite{Quadeer10} gains $0.7$~dB and $0.3$~dB, respectively. 
	The DCCN-CP receiver utilizes CP slightly more effective than the conventional approach \cite{Quadeer10}.

	\subsection{Generalizability}\label{sec:eval:gen}
	
	\begin{table}[!t]
		\renewcommand{\arraystretch}{1}
		\caption{Alternative flow-graphs of DCCN basic receiver with test results} 
		\vspace{-0.1in}
		\label{tab:cdnn:alt}
		\centering
		\footnotesize
		\begin{tabular}{l|c|c|c|c|c|c|c}
			Layers & original &  a &  b &  c &  d &  e &  f   \\ \hline\hline
			DFT-Like C-Conv  & -- & dense & -- & -- & -- & -- & --  \\ \hline
			Conv $2^m\times1\times2$ & X & -- & -- & X & X & -- & --    \\ \hline
			Conv $2^m\times1\times2^m$ & X & -- & -- & X & X & -- & --   \\ \hline
			A0: Leaky ReLU & -- & --  & -- & X & -- & -- & --  \\ \hline
			Concat(IQ, A0) & -- & --  & -- & IQ & A0 & -- & A0  \\ \hline			
			A1: Leaky ReLU & -- & --  & -- & -- & -- & X & --  \\ \hline\hline 
			AWGN: BPSK & \checkmark & \checkmark  & \checkmark & \checkmark & \checkmark & \checkmark & \checkmark  \\ \hline
			AWGN: QPSK & \checkmark & \checkmark  & \checkmark & \checkmark & \checkmark & \checkmark & \checkmark  \\ \hline
			AWGN: 8QAM & \checkmark & \checkmark & \checkmark  & F & F & \checkmark & \checkmark  \\ \hline
			AWGN: 16QAM & \checkmark & \checkmark  & \checkmark & F & F & \checkmark & \checkmark  \\ \hline
		\end{tabular}
	\end{table}
	
	\begin{figure}[!t]
		\centering
		\includegraphics[width=\linewidth]{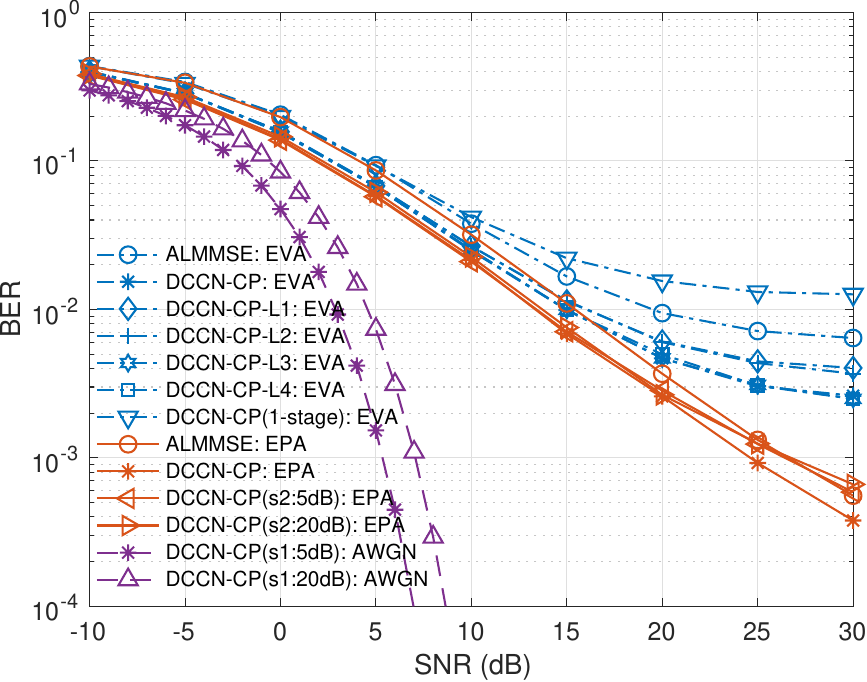}\label{fig:ber:awgn:64}
		\caption{BER performance of DCCN-CP receivers (short CP) with different hyperparameters or training SNRs in EPA and EVA channels. }\vspace{-0.1in}
		\label{fig:ber:alt}
	\end{figure}

	The evaluation of DCCN and benchmarks in different channels in Section~\ref{sec:eval:fading} already shows that a trained equalized DCCN receiver generalizes well to different fading channels. 
	To further illustrate the impact of different training settings on the generalizability of DCCN, we present the BER of a DCCN-CP receiver trained in mixed Rayleigh fading channels in all four different fading channels and an unseen fading channel with $L=12$, in comparison with four other DCCN-CP receivers trained in single Rayleigh fading channel and then tested in the same fading model keeping everything else the same, as illustrated in Fig.~\ref{fig:ber:gen}. 
	When SNR $\geq5$~dB, the receiver trained on mixed fading channels outperforms those trained in single multipath fading channel, except flat fading in which the DCCN-CP receiver trained in mixed fading models underperforms the one trained only in flat fading by $0.5$~dB on average. 
	Mixed fading models could help DCCN overcome local minima during training, as apposed to those trained in single fading model with large delay spread, as shown in Fig.~\ref{fig:ber:gen}, at the cost of precision in flat fading. 
	The advantage of DCCN-CP receiver trained on mixed fading can also be generalized to unseen PDP (green curves in Fig.~\ref{fig:ber:gen}).

	\subsection{Ablation Studies and Alternative Training Methods}\label{sec:eval:alt}
	
	We test 6 alternative structures of the DCCN basic receiver, ($a$--$f$), with some of the 6 components being changed. 
	Their detailed structures and test results for 4 modulations in AWGN channel as listed in Table \ref{tab:cdnn:alt} (--: included, X: removed, \checkmark: test pass, F: test fail). 
	The two consecutive Conv layers are located right before Leaky ReLU $A0$. 
	Except $c$ and $d$, the other 4 alternatives perform identically to the original one.
	Flow-graph $a$ shows that the DFT-Like C-Conv layer can be replaced by a dense layer $\mathbb{C}_{S\times N}$. 
	Large error floor appears in flow-graph $c$ and $d$, which shows that only IQ vector ($c$) or its Leaky ReLU-activation ($d$) can not replace their combination, which, however, can be replaced by linear convolution that expands IQ data from $\mathbb{R}^2$ to $\mathbb{R}^{2^m}$ followed by a non-linear activation in $f$. 
	Alternative $e$ shows that the last Leaky ReLU activation is unessential despite it improves training. 
	Non-linear activation $A0$ helps to reduce the amount of trainable parameters as in the original flow-graph.
	
	Next, we modify the number of dense layers of $\mathbb{C}_{FN\times FN}$ in DCCN-CP equalizers from $0$ (DCCN-CP-L1) to $3$ (DCCN-CP-L4), and remove the 2D C-Conv layer.
	DCCN-CP-L1 and DCCN-CP-L2 underperforms those with $2$ and $3$ layers (L3, L4, and the original) in EVA channels, as the blue lines in Fig.~\ref{fig:ber:alt}, while have similar performance in EPA channels. 
	It shows that two dense layers of $\mathbb{C}_{FN\times FN}$ as in the original one is best for performance and complexity. 
	{It is also verified that by replacing C-Conv layers with an over-simplified implementation of processing real and imaginary parts with two separated Conv layers, DCCN could not be successfully trained, i.e., with a BER of $0.47$.}

	\begin{figure}[!t]
		\centering
		\includegraphics[width=0.95\linewidth]{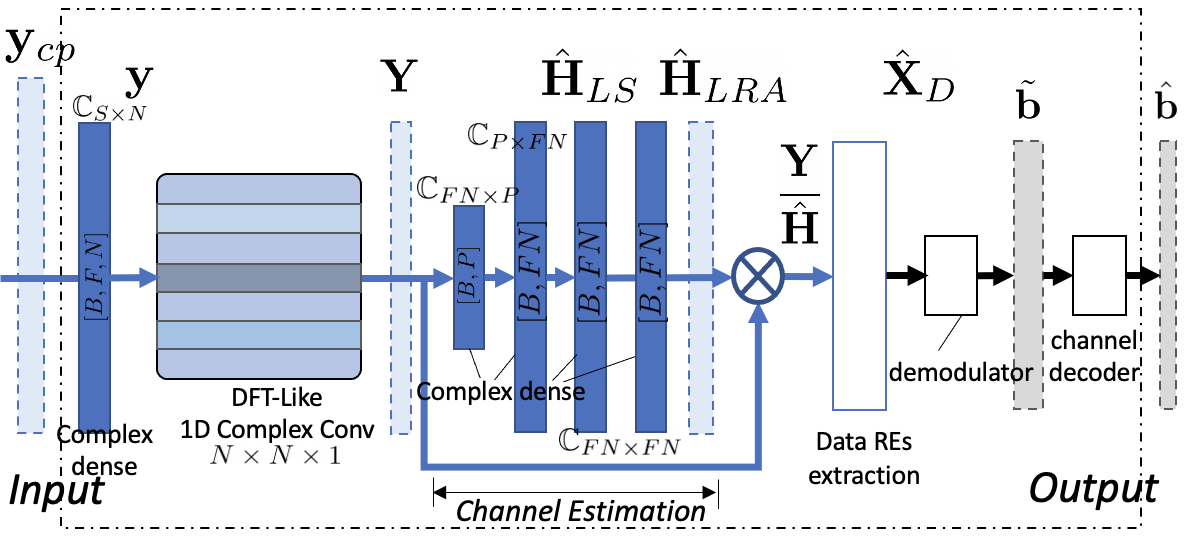}
		\vspace{-0.1in} 
		\caption{Simplified DCCN-CP receiver for deployment, using legacy data REs extraction and demodulation, trained in single stage (stage 2) .}\vspace{-0.1in}
		\label{fig:dccn:alt}
	\end{figure}

	{We also test different training methods for both stages. 
	In stage 1, DCCN-CP trained with an SNR of $20$~dB underperforms the baseline by $2$~dB in AWGN channels as shown by the purple dashed lines in Fig.~\ref{fig:dccn:alt}.} 
	In stage 2, the same initial model trained in mixed fading channels with SNRs of $5$ and $20$~dB, respectively, exhibit larger error floors (0.0003) than the baseline trained in mixed SNRs, as shown by the orange lines in Fig.~\ref{fig:dccn:alt}. 
	{Finally, training the DCCN-CP in a single stage, by mixing the channels and SNRs of the two stages, can significantly degrade the performance, i.e., BER increases by 0.01, as illustrated by the line of \textit{DCCN-CP (1-stage)} in Fig.~\ref{fig:dccn:alt} in EVA channel. 
	These results justify our 2-stage training approach and settings.}

	For deployment, the DCCN-CP receiver structure in Fig.~\ref{fig:flowgraph} can be further simplified by replacing DCCN basic receiver with legacy data extraction and demodulation, as illustrated in Fig.~\ref{fig:dccn:alt}. 
	Compared to \cite{HYe18}, the simplified DCCN-CP receiver requires no explicit DFT, can exploit CP for performance.

	\section{{Conclusion and Future Work}}\label{sec:dis}
	In this paper, an end-to-end OFDM receiver, Deep Complex-valued Convolutional Networks (DCCN), is developed to recover uncoded bits from synchronized time-domain OFDM signal. 
	By following the rule of multiplication in complex field instead of treating the real and imaginary parts of IQ samples as separated streams, DCCN is able to replace DFT/IDFT in OFDM system and exploit the redundancy of cyclic prefix in OFDM waveform for increased SNR. 
	With the expressive power and synergistic advantage of complex-valued neural networks, DCCN is able to combine the tasks of CP-exploitation, low-rank approximation of LMMSE, and inter-symbol interference mitigation, thus outperform the legacy receivers with LMMSE and conventional CP-enhanced channel estimation in doubly-selective Rayleigh fading channels, with a lower computational complexity of $\ccalO(N^2)$.
	This work also offers transferable experience for similar work. 
	Practical guidelines is provided for approximated implementation of complex-valued convolutional networks, especially on setting the dimensions of a convolutional layer with respect to the parameters of OFDM system.
	A suite of novel training methods are developed for deep learning-based wireless transceiver, including a transfer learning scheme, an end-to-end loss function that can prevent vanishing gradient problem in training, and use of mixed SNR and fading models to smooth the loss landscape. 
	It demonstrates the capability of deep neural network in processing sophisticated communication waveform, and suggests that the FFT processor in OFDM receiver can be replaced by a hardware AI accelerator.
	
	Possible future directions with regard to complex-valued neural networks include: 1) explore non-linear signal processing for better performance and/or complexity, 2) improve the scalability by using convolutional layers for channel estimation, 3) extend to spatial domain such as massive MIMO, and 4) explore waveform design and channel coding through complex-valued neural networks-based communication autoencoder.

	
	\bibliographystyle{IEEEtran}
	\bibliography{CogTV,MLRF}

\vspace{-0.1in}
\begin{IEEEbiography}
	[{\includegraphics[width=1in,height=1.25in,clip,keepaspectratio]{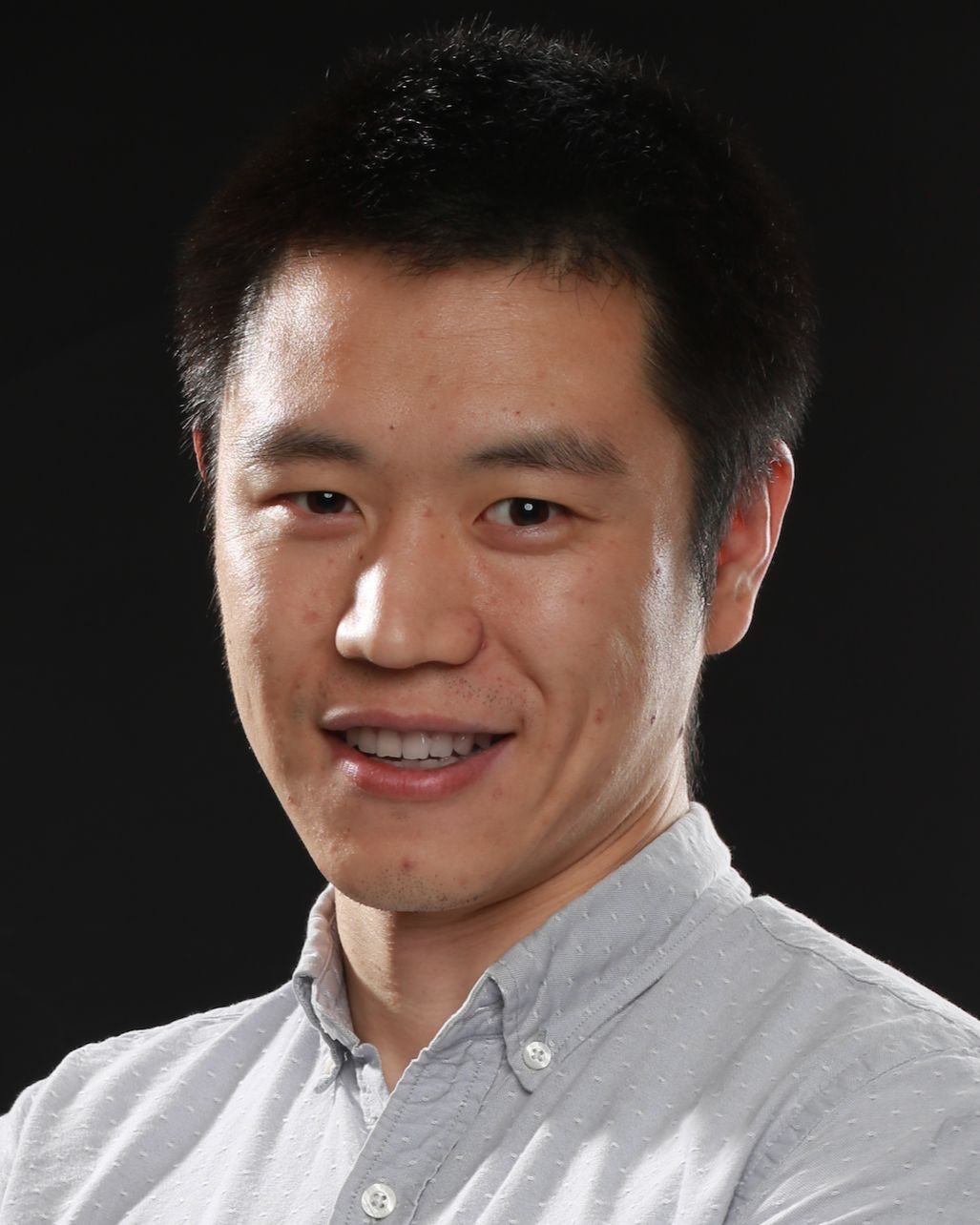}}]{Zhongyuan Zhao} (S'13--M'18) received his B.Sc. and M.S. degrees in Electronic Engineering from the University of Electronic Science and Technology of China, Chengdu, China, in 2006 and 2009, respectively. He received his Ph.D. degree in Computer Engineering from the University of Nebraska-Lincoln, Lincoln, NE, in 2019, under the guidance of Prof. Mehmet C. Vuran. From 2009 to 2013, he worked for ArrayComm and Ericsson, respectively, as an engineer developing 4G base-station. Currently, he is a postdoctoral research associate at the Department of Electrical and Computer Engineering of Rice University. Dr. Zhao’s current research interests focus on machine learning for wireless networks.
\end{IEEEbiography}
\vspace{-0.4in}
\begin{IEEEbiography}
	[{\includegraphics[width=1in,height=1.25in,clip,keepaspectratio]{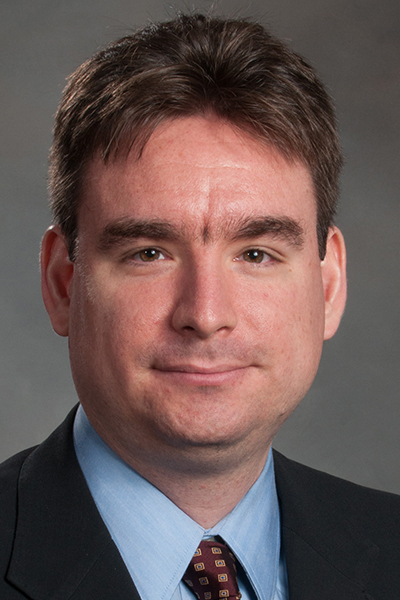}}]{Mehmet Can Vuran} (S'98--M'07) was born in Istanbul, Turkey. He received his B.Sc. degree in Electrical and Electronics Engineering from Bilkent University , Ankara, Turkey in 2002. He received his M.S. and Ph.D. degrees in Electrical and Computer Engineering from the Broadband and Wireless Networking Laboratory, School of Electrical and Computer Engineering, Georgia Institute of Technology, Atlanta, GA. in 2004 and 2007, respectively, under the guidance of Prof. Ian F. Akyildiz. Currently, he is the Susan J. Rosowski Professor of Computer Science and Engineering at the University of Nebraska-Lincoln. Dr. Vuran was awarded an NSF CAREER award for the project ``Bringing Wireless Sensor Networks Underground''. He is a Daugherty Water of Food Institute Fellow.
\end{IEEEbiography}
\vspace{-0.4in}
\begin{IEEEbiography}
	[{\includegraphics[width=1in,height=1.25in,clip,keepaspectratio]{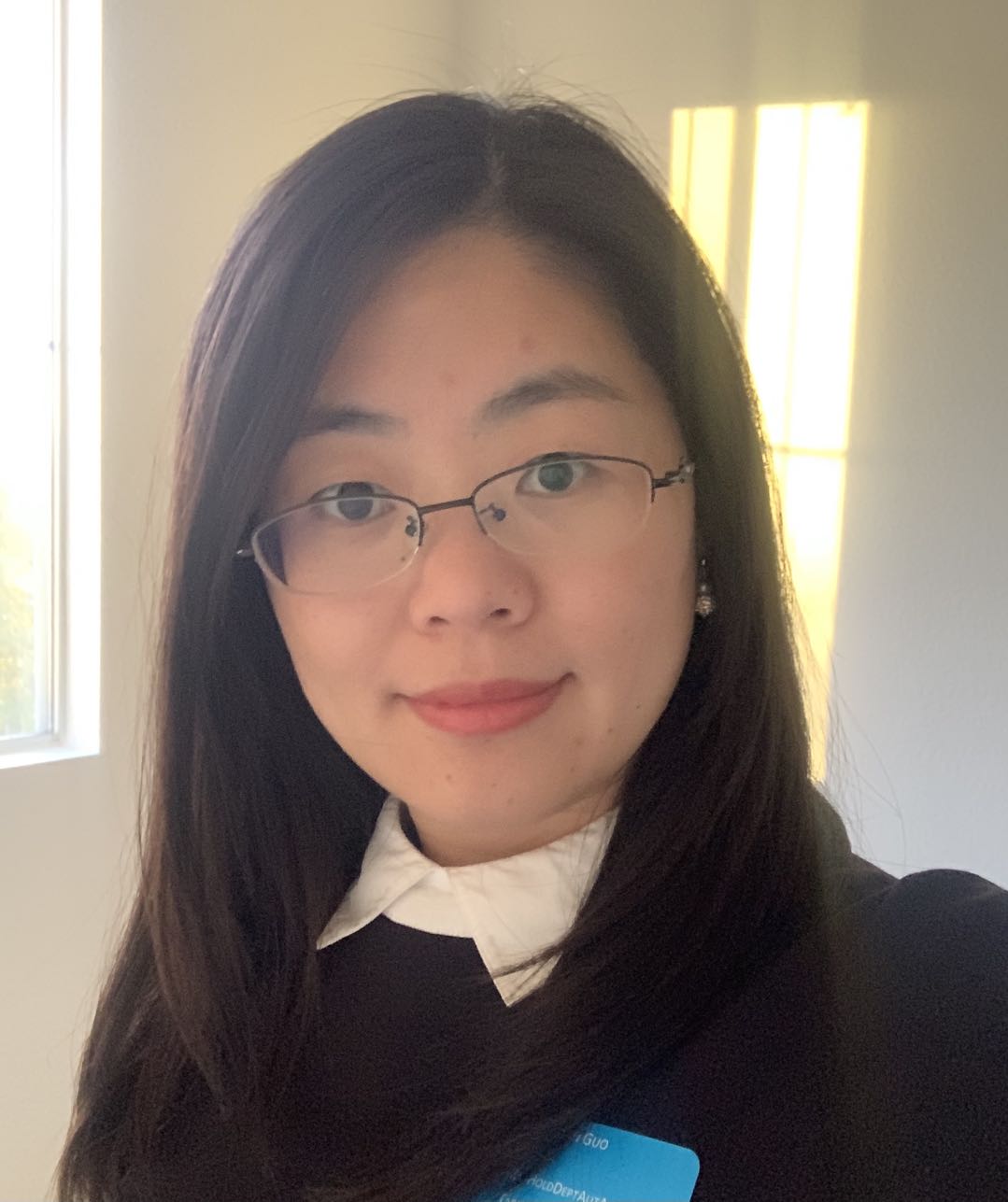}}]{Fujuan Guo} (S'17) received her Ph.D. degree in Computer Engineering from the University of Nebraska-Lincoln, Lincoln, NE, in 2019, under the guidance of Prof. Mehmet C. Vuran. She received her M.S. degree in Electrical Engineering from Harbin Institute of Technology, Harbin, China, in 2012, and B.Sc. degree in Electrical Engineering from Harbin University of Science and Technology, in 2010, respectively. Currently, she is a senior engineer at Qualcomm. Her research interests include wireless communications, channel estimation, time synchronization for the Internet of Things (IoT) system, IoT-based localization, and machine learning in wireless communication. She was a recipient of the Best in Session Presentation Award of IEEE INFOCOM in 2017.
\end{IEEEbiography}
\vspace{-0.4in}
\begin{IEEEbiography}
	[{\includegraphics[width=1in,height=1.25in,clip,keepaspectratio]{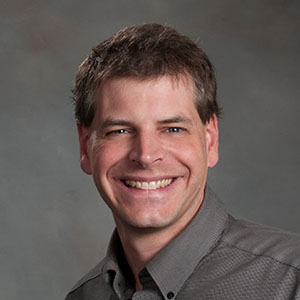}}]{Stephen D.~Scott} received the D.Sc.~degree in Computer Science from Washington University in St.~Louis in 1998. He is  associate professor at the Department of Computer Science and Engineering, University of Nebraska-Lincoln. His current research interests are deep learning and application of machine learning in image and text analysis. 
\end{IEEEbiography}
	
\end{document}